\documentclass[fleqn,usenatbib]{mnras}
\usepackage{newtxtext,newtxmath}

\usepackage[T1]{fontenc}
\usepackage{threeparttable}
\usepackage{xurl}

\usepackage{graphicx}	
\usepackage{amsmath}	
\usepackage{soul}
\usepackage{multirow}
\usepackage[version=4]{mhchem}
\usepackage{amsmath}  
\usepackage{booktabs}

\usepackage{orcidlink}   

\usepackage{booktabs}  

\title[WASP-52\,b with JWST NIRISS/SOSS]{Transmission spectroscopy of WASP-52\,b with JWST NIRISS: Water and helium atmospheric absorption, alongside prominent star-spot crossings}

\author[Marylou Fournier-Tondreau et al.]{Marylou Fournier-Tondreau\orcidlink{0000-0002-5428-0453}$^{1,2}$\thanks{E-mail: marylou.fourniertondreau@physics.ox.ac.uk}, Yanbo Pan\orcidlink{0009-0002-8461-6111}$^{3}$, Kim Morel\orcidlink{0000-0002-1901-1266}$^{2}$, David Lafrenière\orcidlink{0000-0002-6780-4252}$^{2}$, Ryan J. \newauthor{MacDonald\orcidlink{0000-0003-4816-3469}$^{3}$\thanks{NASA Hubble Fellowship Program (NHFP) Sagan Fellow}, Louis-Philippe Coulombe\orcidlink{0000-0002-2195-735X}$^{2}$, Romain Allart\orcidlink{0000-0002-1199-9759}$^{2}$\thanks{SNSF Postdoctoral Fellow}, Loïc Albert\orcidlink{0000-0003-0475-9375}$^{2}$, Michael Radica\orcidlink{0000-0002-3328-1203}$^{4,2}$\thanks{NSERC Postdoctoral Fellow},}\newauthor{Caroline Piaulet-Ghorayeb\orcidlink{0000-0002-2875-917X}$^{4}$, Pierre-Alexis Roy\orcidlink{0000-0001-6809-3520}$^{2}$, Stefan Pelletier\orcidlink{0000-0002-8573-805X}$^{5,2}$, Lisa Dang\orcidlink{0000-0003-4987-6591}$^{2}$\thanks{Banting and Trottier Postdoctoral Fellow}, René Doyon\orcidlink{0000-0001-5485-4675}$^{2}$,}\newauthor{Björn Benneke\orcidlink{0000-0001-5578-1498}$^{2}$, Nicolas B. Cowan\orcidlink{0000-0001-6129-5699}$^{6,7}$, Antoine Darveau-Bernier\orcidlink{0000-0002-7786-0661}$^{2}$, Olivia Lim\orcidlink{0000-0003-4676-0622}$^{2}$, Étienne}\newauthor{Artigau\orcidlink{0000-0003-3506-5667}$^{2,8}$, Doug Johnstone\orcidlink{0000-0002-6773-459X}$^{9,10}$, Lisa Kaltenegger\orcidlink{0000-0002-0436-1802}$^{11}$}, Jake Taylor\orcidlink{0000-0003-4844-9838}$^{1,2}$, Laura Flagg\orcidlink{0000-0001-6362-0571}$^{12}$\\\\
Affiliations are listed at the end of the paper}

\date{Accepted XXX. Received YYY; in original form 2024 December 22}

\pubyear{2024}

\begin{document}
\label{firstpage}
\pagerange{\pageref{firstpage}--\pageref{lastpage}}

\maketitle

\begin{abstract}
In the era of exoplanet studies with JWST, the transiting, hot gas giant WASP-52\,b provides an excellent target for atmospheric characterization through transit spectroscopy. WASP-52\,b orbits an active K-type dwarf recognized for its surface heterogeneities, such as star-spots and faculae, which offers challenges to atmospheric characterization via transmission spectroscopy. Previous transit observations have detected active regions on WASP-52 through crossing events in transit light-curves and via the spectral imprint of unocculted magnetic regions on transmission spectra. Here, we present the first JWST observations of WASP-52\,b. Our JWST NIRISS/SOSS transit observation, obtained through the GTO 1201 Program, detects two clear spot-crossing events that deform the 0.6--2.8\,$\mu$m transit light-curves of WASP-52\,b. We find that these two occulted spots combined cover about 2.4\,\% of the stellar surface and have temperatures about 400--500\,K colder than the stellar photosphere. Our NIRISS/SOSS transmission spectrum is best-fit by an atmosphere with H$_2$O (10.8\,$\sigma$), He (7.3\,$\sigma$, with evidence of an escaping tail at $\sim$\,2.9\,$\sigma$), hints of K (2.5\,$\sigma$), and unocculted star-spots and faculae (3.6\,$\sigma$). The retrieved H$_2$O abundance ($\log$\,H$_2$O $\approx -4 \pm 1$) is consistent with a subsolar or solar atmospheric metallicity for two independent data reductions. Our results underscore the importance of simultaneously modelling planetary atmospheres and unocculted stellar heterogeneities when interpreting transmission spectra of planets orbiting active stars and demonstrate the necessity of considering different stellar contamination models that account for both cold and hot active regions.

\end{abstract}

\begin{keywords}
planets and satellites: atmospheres -- planets and satellites: gaseous planets -- planets and satellites: individual: WASP-52\,b - starspots -- methods: data analysis - techniques: spectroscopic.
\end{keywords}



\section{Introduction}
Transmission spectroscopy has proven to be one of the most powerful techniques for probing the composition of exoplanet atmospheres and making inferences about the formation and migration pathways of exoplanets \citep{seager2000,brown2001b}. The James Webb Space Telescope (JWST) is now the leading observatory for characterizing the atmospheres of exoplanets through transmission spectroscopy \citep[e.g.,][]{fisher2024}. The observations made with this new telescope have been fruitful so far, showing exquisite precision, revealing several spectral features, and leading to the first detections of CO$_2$ and SO$_2$ in the atmosphere of a hot Jupiter \citep[WASP-39\,b;][]{ers2023,rustamkulov2023,alderson2023}. Beyond providing better precision, JWST also improves on the spectral range available to its predecessors and now covers the absorption features of the most abundant molecules expected in hot Jupiter's atmospheres, such as H$_2$O, CH$_4$, CO, and CO$_2$, as well as alkali metals, like Na and K \citep[e.g.,][]{burrows1999}. With sensitivity to such a wide range of molecules, JWST offers an opportunity to study many out-of-equilibrium effects that can occur in exoplanet atmospheres, such as vertical mixing and photochemistry \citep[e.g.,][]{welbanks2024,tsai2023}. Also, JWST can provide new insights into the presence and characteristics of hazes and clouds \citep[e.g.,][]{inglis2024,bell2024}. A wide range of exotic condensates is proposed to exist, ranging from pure transition metals (like Fe) to silicates, sulphides, and salt/alkali condensates, which may give rise to significant cloud coverage \citep[e.g.,][]{wakeford2015}. However, clouds can also complicate transmission spectroscopy analyses by covering considerable portions of the atmosphere, making it challenging to identify the chemical species present therein \citep[e.g.,][]{deming2013}. Stellar contamination has proven to be another main challenge in studying the atmospheres of exoplanets, limiting the inferences for rocky planets \citep[e.g.,][]{lim_atmospheric_2023,moran2023,radica2024b}, and potentially biasing those of inflated, gas giants if not well accounted for \citep[e.g.,][]{barstow2015,fournier-tondreau2024}. 

The inflated ($R$ = 1.27\,$R_J$), Saturn-mass exoplanet ($M$ = 0.46\,$M_J$) WASP-52\,b \citep{hebrard2013} is an ideal target for atmospheric studies given its deep transit ($\delta$ = 2.71\,\%), and large scale height ($H$ $\approx$ 700\,km), which is due to its high temperature ($T_{\rm eq}$ = 1315\,K; orbital period about 1.75 days) and low surface gravity ($\log g_p$ = 2.81\,cm s$^{-2}$). Prior analyses have often found muted spectral features in transmission spectra of WASP-52\,b, which have been attributed to a cloudy atmosphere \citep{kirk2016,chen2017,alam2018}; nonetheless, water vapour absorption has been observed in the near-infrared \citep{tsiaras2018,bruno2018,bruno2020}, whereas sodium, potassium, hydrogen, and helium have been detected at high-resolution \citep{chen2020,kirk2022,canocchi2024}. 

The effect of its young and active K2V host star was carefully considered in past studies observing the planet in transit \citep[e.g.,][]{kirk2016,bruno2020}. Stellar activity has indeed long been acknowledged as a significant obstacle in characterizing exoplanet atmospheres through transmission spectroscopy \citep[e.g.,][]{pont2008,czesla2009}. The presence of stellar heterogeneities, such as star-spots and faculae, can result in bumps or dips in light-curves when situated along the transit path \citep[e.g.,][]{pont2007}. Outside the transit chord, the stellar heterogeneities can cause a mismatch between the light source sampled by the planet's transit and the assumed full stellar disk spectrum, resulting in the so-called transit light source effect (TLSE; \citealp{rackham2018}). Therefore, occulted active regions can hinder the correct measurement of transit parameters and depths by substantially impacting transit light-curves \cite[e.g.,][]{barros2013,oshagh2014}, whereas unocculted active areas can introduce spurious spectral features by the TLSE \citep[e.g.,][]{mccullough2014,rackham2018}. 

Many efforts have been undertaken to mitigate the impact of stellar activity and achieve unbiased atmospheric inferences (see \citealp{rackham2023a} for a recent review with current recommendations). For occulted active regions, simply masking them when fitting light-curves does not completely negate their effect; accounting for them with a spot-transit model is the recommended strategy to mitigate their impact \citep{rackham2023a}; otherwise, these magnetic regions can potentially imprint strong slopes towards short visible wavelength \citep[e.g.,][]{oshagh2014}. Previous studies of WASP-52\,b have revealed transit light-curve anomalies associated with occultations of star-spots \citep{mancini2017,bruno2018,may2018} and a facula \citep{kirk2016}. When modelling active regions, the standard approach is to parametrize their position and size on the stellar surface from a ``white'' light-curve and then infer their temperature with the spectrophotometric light-curves. However, degeneracies can exist between all of those parameters \citep{fournier-tondreau2024,rackham2023a}, not only between the size and temperature \citep[e.g.,][]{pont2008}. One approach worth considering, which we explore here, is the simultaneous and joint fitting of all spectrophotometric light-curves using both wavelength-dependent and -independent parameters. This method circumvents the need to fix the positions and sizes of active regions based on a broadband light-curve fit, thereby better exploiting the wavelength-dependence effect of stellar heterogeneities. 

For unocculted active regions, the current state of the art is to jointly fit for the properties of the stellar heterogeneities and of the planetary atmosphere when performing retrievals on transmission spectra \citep[e.g.,][]{pinhas2018,rathcke2021}. For WASP-52\,b, \citet{bruno2020} presented a joint retrieval analysis of its optical to infrared transmission spectrum combining HST/WFC3 \citep{bruno2018}, HST/STIS and \textit{Spitzer}/IRAC \citep{alam2018} observations. They found unocculted star-spots covering 5\,\% of the stellar surface with a temperature <\,3000 K, and reported a water abundance of $\log$ H$_2$O = $-3.30^{+0.94}_{-1.12}$, a $\sim$\,0.1-10$\times$ solar metallicity and a subsolar carbon-to-oxygen ratio (C/O). Earlier studies \citep[e.g.,][]{alam2018} instead directly corrected the transmission spectrum based on stellar activity monitoring or occulted heterogeneity properties; however, this was proposed to underestimate the impact of unocculted active regions \citep{rackham2023a}. Hot giants can ultimately serve as test cases for validating and improving the mitigation of stellar contamination since they are known to have absorption features with similar amplitudes to those imprinted by stellar heterogeneities, as opposed to terrestrial planets, where TLSE can dominate over atmospheric features. 

Here, we present the 0.6--2.8~$\mu$m transmission spectrum of WASP-52\,b, obtained with the Near Infrared Imager and Slitless Spectrometer (NIRISS) instrument of the JWST. Spot-crossing events deform the transit light-curves. We thus model those active regions in the light-curve fitting process and then jointly fit the properties of the planetary atmosphere and the unocculted stellar heterogeneities by performing retrievals on the resulting transmission spectrum. We briefly outline the observations and data reduction in Section \ref{sec:data}. We describe the light-curve fitting and the modelling of the spot-crossing events in Section \ref{sec:lc}. We detail the retrieval analysis in Section \ref{sec:retrievals} and follow up with the helium absorption analysis in Section \ref{sec:He}. We summarize and discuss the results in Section \ref{sec:discu} and conclude in Section~\ref{sec:conclu}.

\section{Observations \& data reduction} \label{sec:data}
A transit observation of WASP-52\,b was obtained using the Single Object Slitless Spectroscopy (SOSS) mode of the NIRISS instrument \citep{albert2023,doyon2023} as part of the JWST Guaranteed Time Observations (GTO) program cycle 1 (PID: 1201; PI: David Lafrenière). The time series observation (TSO) started on November 27\textsuperscript{th}, 2022, at 07:08:33.169 UTC and spanned 4.44 hours, which covered the 1.8 hr transit as well as 1.9 hr of baseline before and 0.74 hr after the transit. It used the standard GR700XD/CLEAR combination, along with the SUBSTRIP256 detector, which captures diffraction orders 1 and 2 of the SOSS mode. There is a total of 265 integrations, each consisting of 10 groups and lasting 60.434 seconds. 

\begin{table}
\begin{center}
\caption{Parameters of the WASP-52 planetary system.}
\label{tab:1} 
\begin{tabular}{l|cc}\hline\hline
Parameters & WASP-52 & Units\\ \hline
\multicolumn{3}{l}{\textbf{Stellar parameters}}\\\hline
Spectral type   & K2V & \\
Rotational period & 11.8 $\pm$ 3.3 &day\\
Stellar radius & 0.79 $\pm$ 0.02 & R$_\odot$ \\
Effective temperature & 5000 $\pm$ 100& K\\ 
Stellar surface gravity & 4.582 $\pm$ 0.014 & log$_{10}$ cm s$^{-2}$\\
Metallicity & 0.03 $\pm$ 0.12 & [Fe/H]\\ \hline
\multicolumn{3}{l}{\textbf{Planetary and transit parameters}}\\ \hline
Planet radius  & 1.27 $\pm$ 0.03 &R$_{\text{J}}$\\
Planet mass  &0.46 $\pm$ 0.02&M$_{\text{J}}$\\
Planet surface gravity & 2.81 $\pm$ 0.03 &log$_{10}$ cm s$^{-2}$ \\
Orbital period  &1.7497798 $\pm$ 0.0000012& day\\
Orbital eccentricity & 0 & \\
Impact parameter &0.60 $\pm$ 0.02&\\
Scaled semi-major axis & 7.3801 $^{+0.1106}_{-0.1073}$ & \\
Area ratio ($R_{\text{p}}$/$R_*$)$^2$ &0.0271 $\pm$ 0.0004&\\
Transit duration& 0.0754 $\pm$ 0.0005 & day \\
Equilibrium temperature  & 1315 $\pm$ 35 & K\\\hline
\multicolumn{3}{l}{\footnotesize \textit{Note:} Parameter values from \cite{hebrard2013}.}
\end{tabular}
\end{center}
    
\end{table}

We use stages 1 and 2 of the \texttt{exoTEDRF}\footnote{Formerly known as \texttt{supreme-SPOON}. Version 1.1.7 is used in this work.} pipeline \citep[e.g.,][]{radica2023, radica2024exoTEDRF} to reduce the TSO, starting from the raw, uncalibrated files downloaded from the Mikulski Archive for Space Telescopes (MAST). In stage 1, \texttt{exoTEDRF} uses the same steps as the official \texttt{jwst} pipeline, except for the 1/$f$ noise correction, to perform the detector-level calibrations. However, we do not apply the \texttt{RefPixStep} since the 1/$f$ noise correction from \texttt{exoTEDRF} has the same purpose as the \texttt{RefPixStep} of the \texttt{jwst} pipeline, including the correction of the even-odd row variations \citep[e.g.,][]{feinstein2023}.

As explained in \citet{radica2023}, the zodiacal background must be subtracted before correcting for the 1/$f$ noise. Since a constant scaling of the Space Telescope Science Institute (STScI)\footnote{\url{https://jwst-docs.stsci.edu/known-issues-with-jwst-data/niriss-known-issues/niriss-soss-known-issues\#gsc.tab=0}} SOSS SUBSTRIP256 background model does not completely remove the background \citep[e.g.,][]{lim_atmospheric_2023, fournier-tondreau2024}, we separately scale both sides of the pick-off mirror jump of the STScI model. This was done by using regions on the top of the detector above the third order trace, redward ($x \in [228,250],~ y \in [380,574]$) and blueward ($x \in [227,250],~ y \in [793,897]$) of the background jump. The scaling factors between the median frame for each group and the background model were calculated by considering the 16$^{\text{th}}$ and 12$^{\text{th}}$ percentiles of the distribution of the ratios for the left and right side of the background step, respectively, to account for the small amount of flux present in the two regions.

To remove the 1/$f$ noise, which is treated at the group level, we carefully mask every order 0 contaminant as well as the two dispersed contaminants (one in the upper left corner and one below the first order's trace in the center) on the detector to make sure they do not bias the reduction. Once the 1/$f$ correction is done (refer to \citet{radica2023} for a detailed description of this step), the previously subtracted background is re-added to each group in each integration. The final removal of the zodiacal background is performed in stage 2, following the same procedure described above, along with further calibrations such as flat fielding and warm-pixel interpolation.

The 1D spectral extraction was performed using the \texttt{ATOCA} algorithm \citep{darveau-bernier2022}, which takes into account the contamination between the first and second order on the detector. The \texttt{APPLESOSS} code \citep{radica2022} was used to create the \texttt{specprofile} reference file needed for \texttt{ATOCA}. During the extraction, all pixels flagged as \texttt{DO\_NOT\_USE} and all order 0 contaminants are modelled by the \texttt{ATOCA} algorithm. The spectra were extracted on the decontaminated traces using a box width of 30 pixels. We then used the \texttt{PASTASOSS} package\footnote{Available here: \url{https://github.com/spacetelescope/pastasoss}} \citep{Baines2023} to obtain the wavelength solution for WASP-52\,b's observation. Lastly, any data point deviating by more than 5\,$\sigma$ in time was clipped.

A summary of the major reduction steps can be visualized in Figure \ref{fig:Reduction Steps}. Furthermore, we reduced the observations with another independent pipeline, \texttt{NAMELESS} \citep[e.g.,][Coulombe et al. in press]{coulombe2022,radica2023}, to verify the consistency of our results (see Appendix \ref{sec:pipeline}).

\section{Light-curve fitting \& occulted star-spot analysis}\label{sec:lc}
\subsection{White light-curve fitting}
\begin{figure}
	\centering
	\includegraphics[width=\columnwidth]{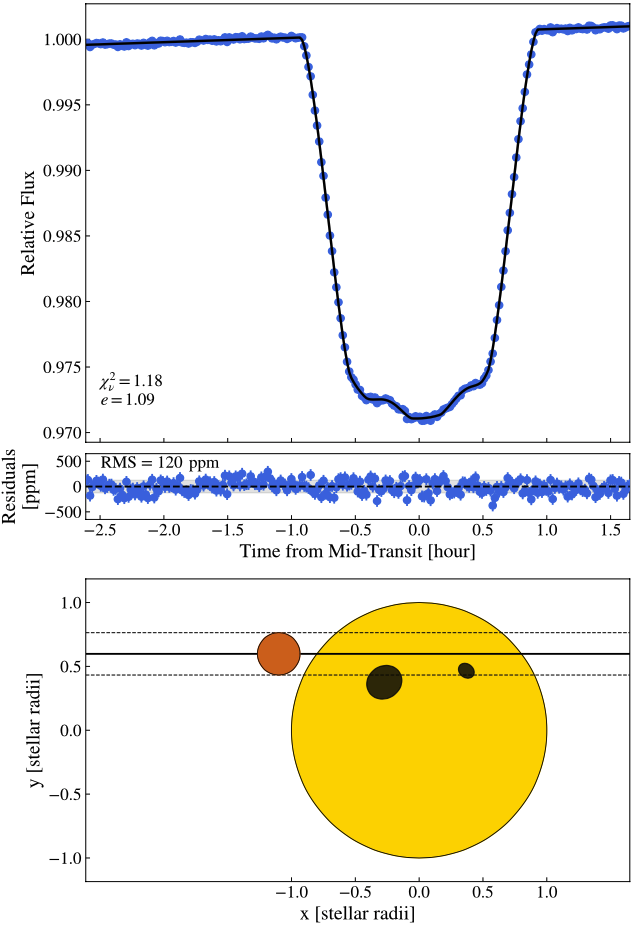}
    \caption{Planetary transit and spot-crossing modelling. \emph{Top}: Broadband light-curve (blue) along with the best-fitting spot-transit model with the highest likelihood overplotted (black). The fit statistics are listed in the bottom left corner (reduced chi-squared $\chi^2_{\nu}$ and error multiple needed to obtain a $\chi^2_{\nu}$ equal to unity \emph{e}). \emph{Middle}: Residuals to the transit fit with the root-mean-square (RMS) scatter. \emph{Bottom}: Physical representation of the solution for the occulted star-spots (black circle) on the star (yellow circle), along with the transit motion in black (dashed lines representing the transit chord) of the planet (orange circle). The system, including the star-spots, is up-to-scale. 
    \label{fig:starspot_lc}}
\end{figure}

We construct a broadband light-curve, following \citet{fournier-tondreau2024}, by summing the flux from wavelengths bluewards of 1.5\,$\mu$m in order 1 (0.85–1.5\,$\mu$m) and from 0.65–0.85\,$\mu$m in order 2 to keep the wavelength range where the spot-crossings have a more substantial effect. We mask the first 10 integrations and the 65\textsuperscript{th}, which had an anomalous background signal. The resulting broadband light-curve is shown in Figure \ref{fig:starspot_lc} and displays two clear spot-crossing events, seen as bumps near the beginning and end of the transit. We fit a spot-transit model with two spot-crossing events using \texttt{spotrod} \citep{beky2014} with the \texttt{Juliet} package \citep{espinoza2019}. We fix the orbital period to 1.7497798 d and the eccentricity to 0 (\citealp{hebrard2013}; all the values used in this paper are listed in Table \ref{tab:1}). We fit for the mid-transit time $t_0$, the impact parameter $b$, the scaled semi-major axis $a/R_*$, the scaled planet radius $R_\mathrm{p}/R_*$, the spots' $x$- and $y$-position, the spots' radius $R_\mathrm{spot}$, the spot-to-stellar flux contrast $F_\mathrm{spot}/F_*$, a term to fix the zero point of the transit baseline $\rm \theta_0$, and the two quadratic limb darkening (LD) parameters ($q_1$, $q_2$) following the parameterization of \citet{kipping2013}. We also fit for a scalar jitter term, $\sigma$, which is added in quadrature to the flux error. We tested detrending against linear models with time, trace $x$-position, trace $y$-position, and a linear and quadratic model with time. We find that the broadband light-curve is best fit by a transit model with a slope with time $\rm \theta_1$. We fit 17 parameters and sample the parameter space with 2000 live points using \texttt{dynesty} \citep{speagle2020}. The priors and the best-fitting transit and spot parameters for the broadband light-curve fit are shown in Table \ref{tab:wlc_parameters}. The reduced chi-squared statistic for the fit with the highest likelihood is $\chi^2_\nu$ = 1.18. This best-fitting spot-transit model is overplotted in the top panel of Figure \ref{fig:starspot_lc}, and a physical representation of the spot-crossings is shown in the bottom panel. We also tested a model with a facula (bright region) near the middle of the transit instead of two dark spots by allowing the contrast to be greater than one, but we ruled it out because the model could not reproduce the sharpness of the two bumps nor their asymmetry. Furthermore, we assessed the H-alpha light-curve and did not detect any features that could be indicative of flares.

\begin{figure}
	\centering
	\includegraphics[width=\columnwidth]{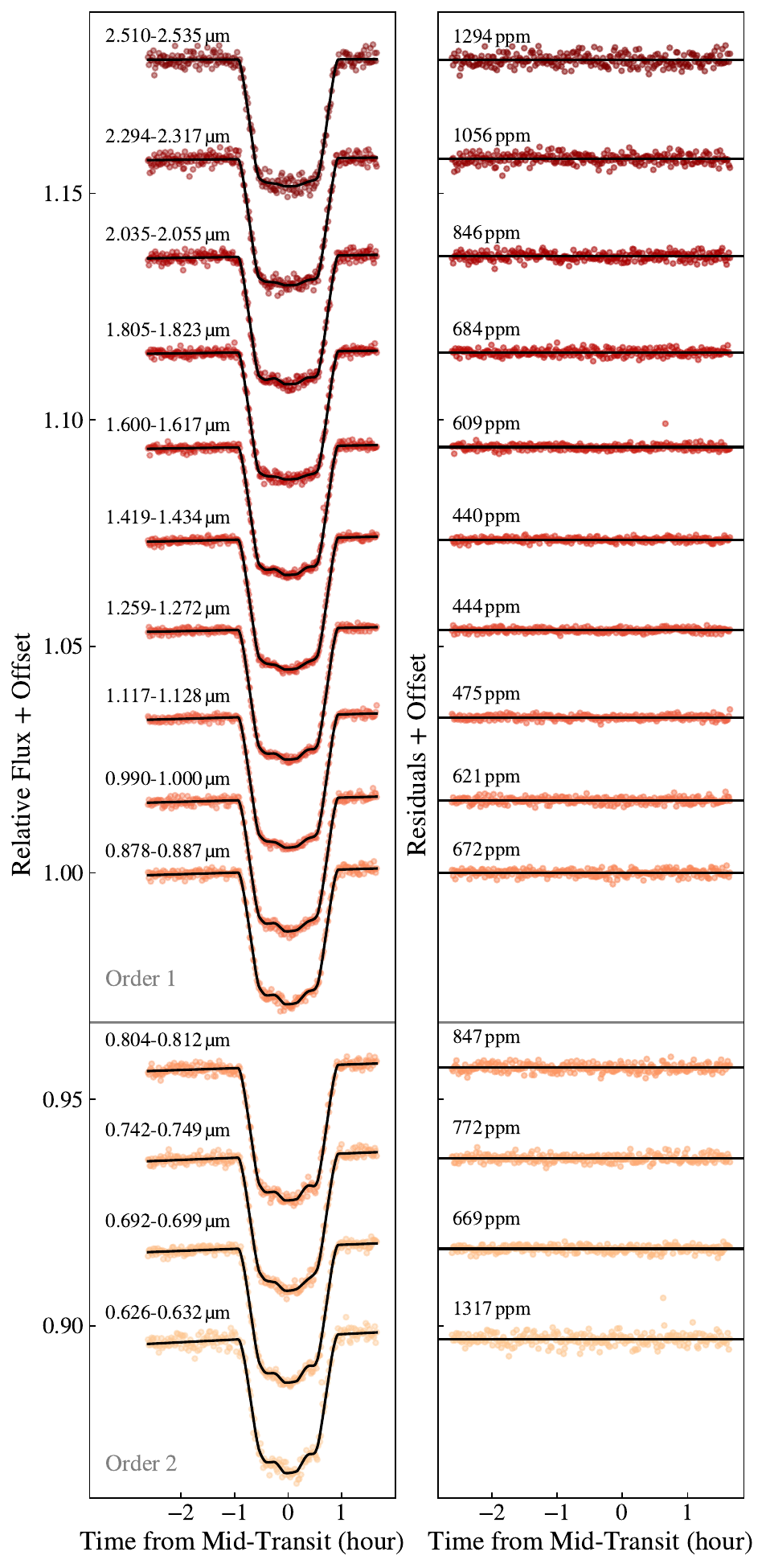}
    \caption{Normalized spectrophotometric light-curves at a resolving power of $R$ = 100. \emph{Left}: Binned spectrophotometric light-curves, along with the best-fitting spot-transit models for different spectral bins (black). \emph{Right}: Associated residuals to the transit fit in each bin with the RMS scatter indicated. 
    \label{fig:slc}}
\end{figure}

\subsection{Spectrophotometric light-curve fitting}

\begin{figure*}
	\centering
	\includegraphics[width=\textwidth]{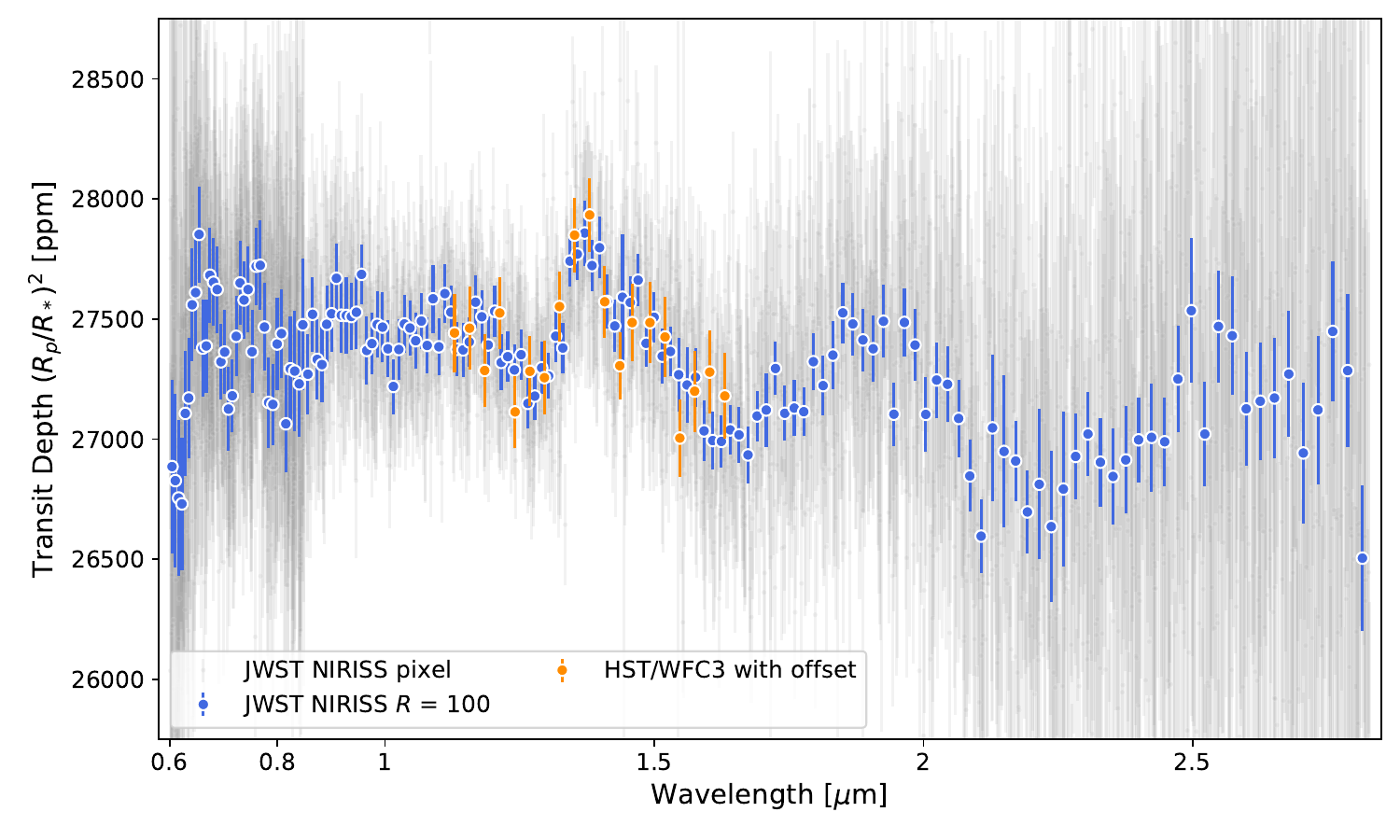}
    \caption{JWST NIRISS transmission spectrum of WASP-52\,b at pixel resolution (faded grey) and binned to a resolving power of $R$ = 100 (blue). The HST/WFC3 transmission spectrum retrieved by \citet{bruno2018} with a spot-transit model is shown for comparison (orange); note that an offset of -250 ppm has been applied. Bar this overall offset, the shape of NIRISS and WFC3 transmission spectra show a remarkable agreement where they overlap in wavelength.}
    \label{fig:transmission_spectrum}
\end{figure*}

We proceed to fit the spectrophotometric light-curves both at the pixel level and at a resolving power of $R$ = 100. At this point, $t_0$, $b$, $a/R_*$, the position and radius of the spots were fixed to the best-fitting values with the highest likelihood from the broadband fit following \citet{fournier-tondreau2024}. The remaining transit parameters to be fitted were the scaled planet radius, the contrast for each spot, the two quadratic LD parameters for each spectrophotometric light-curve, alongside $\sigma$, $\rm \theta_0$ and $\rm \theta_1$. We put Gaussian priors on the LD parameters based on calculations from the \texttt{ExoTiC-LD} package \citep{wakeford2022} using the 3D stagger grid \citep{magic2015}. The widths of the Gaussian priors are set to 0.2 following \citet{patel2022}. We use 500 live points for each spectral bin. The spectrophotometric light-curves for 14 bins at a resolving power of $R$ = 100, along with their corresponding best-fitting spot-transit model, are displayed in Figure \ref{fig:slc}. The resulting transmission spectra at the pixel level and $R$ = 100 are in agreement, as shown in Figure \ref{fig:transmission_spectrum}. It was better to fit the spectrophotometric light-curves at this lower resolution because the model cannot constrain the spot contrasts properly at the pixel level at wavelengths redwards of 1.8\,$\mu$m due to the low signal-to-noise ratio, but that effect does not impact the transit depths. The transmission spectrum at $R$ = 100 was used for the retrieval analysis. 

\subsection{Inferred occulted star-spot properties on WASP-52}
\begin{figure*}
    \centering
 \includegraphics[width=\textwidth]{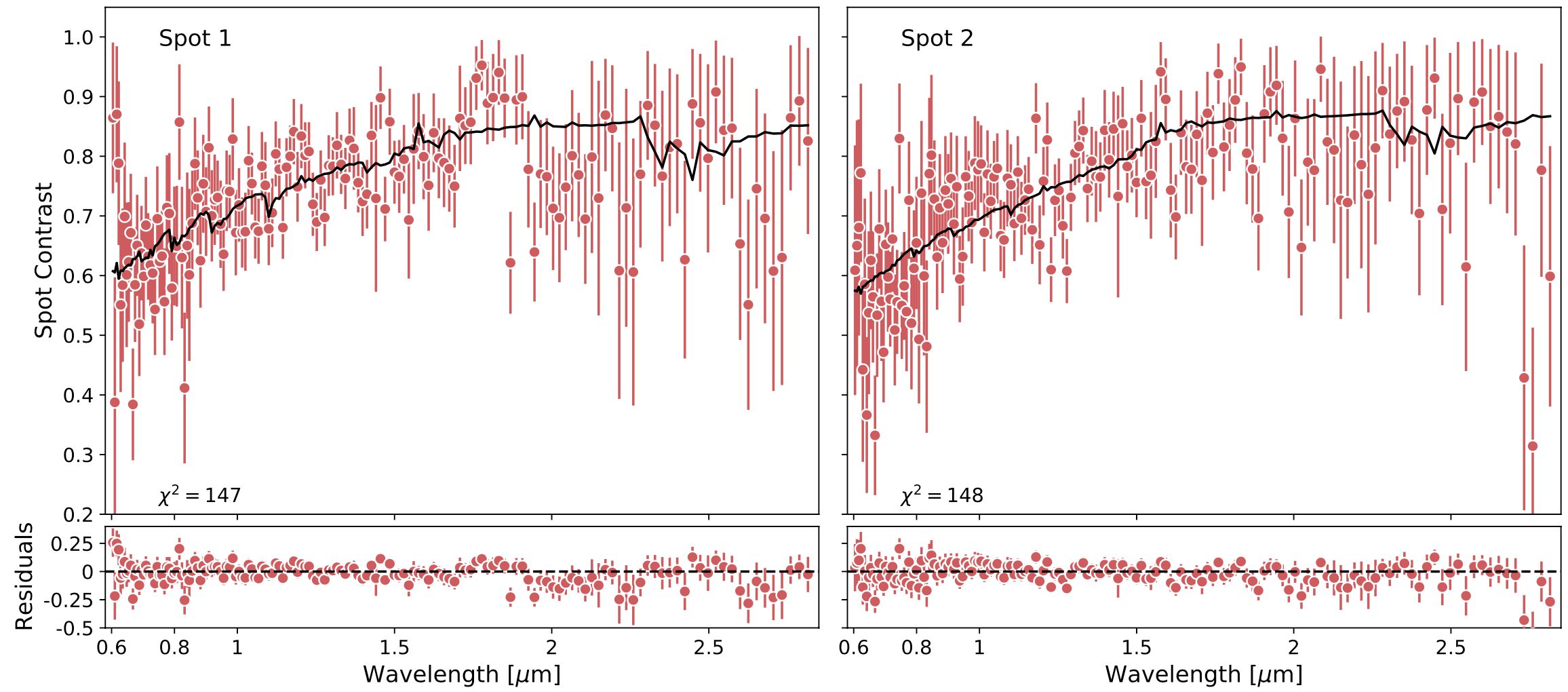}
    \caption{Contrast spectra of the two star-spot crossings; the one occulted before the mid-transit (left panel) and the one after (right panel). \emph{Top}: Retrieved spot contrast spectrum at a resolving power of $R$ = 100 (blue), along with the overplotted best-fitting contrast model for each spot (black). The chi-squared ($\chi^2$) is listed in each bottom left corner. \emph{Bottom}: Residuals to the contrast fit for each spot.
    \label{fig:contrast}}
\end{figure*}

We constrain the temperature of each occulted star-spot by fitting PHOENIX synthetic stellar spectra \citep{husser2013} to the retrieved contrast spectrum of each spot. We fit for the spot temperature and set the surface gravity of the spot model as a free parameter following \citet{fournier-tondreau2024}. We model each spot contrast spectrum by taking the flux ratio of a spot spectrum to the star spectrum. For the spot model, we use PHOENIX stellar models with temperatures from 4000 to 5000\,K, logarithmic surface gravities ($\log\,g$) from 1.5 to 5.5\,dex and a fixed metallicity of 0.03 from \citet{hebrard2013}, and we interpolate these spectra linearly in temperature and $\log\,g$. We compute a stellar spectrum for the star model with a temperature, $\log g$, and metallicity fixed to \citet{hebrard2013} values.  We employ \texttt{dynesty} \citep{speagle2020} with 500 live points to chart the parameter space. Figure \ref{fig:contrast} shows for each occulted spot the best-fitting contrast model overplotted on each retrieved spot contrast spectrum.

The broadband light-curve fit led to precise measurements of the position of the spot occulted after the mid-transit but not to a well-constrained position for the first spot. Similarly to \citet{fournier-tondreau2024}, the best-fitting value of the $y$-position varies substantially between different broadband light-curve fits (without changing anything), and it is highly correlated with its size and contrast (see corner plot in Figure \ref{fig:corner_wlc}). Nonetheless, we find two cold spots of sizes $R\rm _{spot,1}$ = 0.13$^{+0.05}_{-0.04}$\,R\textsubscript{*} and $R\rm _{spot,2}$  = 0.085$^{+0.03}_{-0.019}$\,R\textsubscript{*}, with corresponding temperatures of $\rm \Delta T_1$ = 420 $\pm$ 20\,K and $\rm \Delta T_2$ = 480 $\pm$ 20\,K colder than the photosphere. The coverage fraction of these two occulted spots combined is thus about 2.4\,\% of the visible stellar hemisphere. The surface gravities of the spot models are lower than the star by $\Delta \log g_1$ = 2.4$^{+0.4}_{-0.3}$\,dex and by $\Delta \log g_2$ = 1.4$^{+0.6}_{-0.5}$\,dex. These differences on the $\log g$ for the spot model are similar to what we infer in \citet{fournier-tondreau2024} but bigger than physically expected. The increased magnetic pressure in active features decreases the gas pressure, which can be represented by a stellar model with a 0.5--1\,dex lower surface gravity \citep{solanki2003,bruno2022}. Therefore, we tested fitting only for the spot temperature, fixing the surface gravity to the stellar value. We find for the second occulted spot, the one that is well-constrained, that there is only weak evidence (log Bayes factor $\ln \mathcal{B}_{01}$ = 1.3) to set the surface gravity of the spot model as a free parameter, whereas for the first spot, there is strong evidence ($\ln \mathcal{B}_{01}$ = 6.3). Still, that does not impact the retrieved spot temperatures significantly. 

Furthermore, we also explored another light-curve fitting method that does not rely on a broadband light-curve fit in an attempt to lift some degeneracies. In this approach, all spectral light-curves are simultaneously and jointly fit by treating some parameters common to all spectral bins (achromatic) and some parameters varying for each spectral bin (chromatic). With this new approach, we retrieve the same solution for the second occulted star-spot and still find degenerate solutions for the $y$-position of the first one, where the $y$-position is orthogonal to the path of the planet's transit, leading to different spot sizes and temperatures.

\section{Retrieval analysis} \label{sec:retrievals}

We present inferences from a Bayesian retrieval analysis applied to WASP-52\,b's transmission spectrum. Given the strong signature of occulted star-spots in the spectrophotometric light-curves, we additionally consider here the influence of unocculted stellar active regions jointly with atmospheric models. In what follows, we first outline our retrieval configuration before detailing our interpretation of WASP-52\,b's transmission spectrum in the light of planetary atmosphere absorption and stellar contamination. 

\subsection{Retrieval configuration} \label{subsec:retrieval_config}

We perform retrievals on WASP-52\,b's transmission spectrum using the open source retrieval code \textsc{Poseidon} \citep{macDonald2017,macDonald2023}, which uses the MultiNest algorithm to explore the multidimensional parameter space. All retrievals are conducted with 1,000 MultiNest live points to explore the posterior distributions of the parameter space smoothly.

Our atmospheric model assumes an isothermal pressure-temperature (P-T) profile dominated by \ce{H2} and \ce{He} (with an assumed ratio of \ce{He}/\ce{H2} = 0.17). The atmospheric model spans $10^{-7}$--10$^2$\,bar, with 100 layers spaced uniformly in log pressure, and uses a reference pressure of 10\,bar as the boundary condition for hydrostatic equilibrium (i.e., the pressure where the retrieved reference radius is located). We consider several trace chemical species expected at the equilibrium temperature of WASP-52\,b \citep{Madhusudhan2016,Woitke2018,Mukherjee2024} with strong absorption features in the NIRISS/SOSS wavelength range: \ce{H2O}, \ce{Na}, \ce{K}, \ce{CO}, \ce{CO2}, \ce{CH4}, \ce{NH3} and \ce{HCN}. Our opacities use state-of-the-art line lists and pressure broadening parameters included in the recent \textsc{Poseidon} v1.2 \citep{Mullens2024} release, from the following line list sources: \ce{H2O} \citep{Polyansky2018}, \ce{Na} and \ce{K} \citep{Ryabchikova2015},  \ce{CO} \citep{li2015}, \ce{CO2} \citep{Yurchenko2020}, \ce{CH4} \citep{Yurchenko2024}, \ce{NH3} \citep{Coles2019}, and \ce{HCN} \citep{Barber2014}. We also include collision-induced absorption from \ce{H2}--\ce{H2} and \ce{H2}--\ce{He} \citep{Karman2019} and \ce{H2} Rayleigh scattering \citep{Hohm1994}. Finally, we follow a parametric treatment of aerosols via the 4-parameter inhomogeneous cloud and haze prescription from \citet{macDonald2017}.  

Given the evidence of unocculted stellar features from the HST transmission spectra of WASP-52\,b \citep{bruno2020}, we also include stellar contamination parameters in our retrievals. We use three sets of model configurations, as described in \citet{fournier-tondreau2024}: (i) atmosphere-only, (ii) one-heterogeneity + atmosphere, and (iii) star-spots + faculae + atmosphere. We applied all three models to the \texttt{exoTEDRF} transmission spectrum. We also explore the sensitivity of data reduction by running atmosphere-only and star-spots + faculae + atmosphere retrievals on the \texttt{NAMELESS} spectrum. The one heterogeneity model is defined by the stellar photosphere temperature, the heterogeneity temperature (less than the photosphere for spots, greater than the photosphere for faculae), and the heterogeneity covering fraction. The star-spots + faculae model has two heterogeneities, one assumed colder than the photosphere (spot), and one assumed warmer than the photosphere (faculae), each with their own covering fraction. Thus, the one-heterogeneity model adds three free parameters, while the star-spots + faculae model adds five free parameters. We additionally experimented with retrieving different surface gravities for the heterogeneities compared to the photosphere (as in \citealt{fournier-tondreau2024}) but found this unnecessary. We calculate the contribution from stellar contamination by interpolating PHOENIX models \citep{husser2013} using the PyMSG package \citep{Townsend2023}.

We calculate model spectra at a spectral resolution of $R = \lambda/d\lambda = 20,000$ from 0.58 to 2.84\,$\mu$m using the configuration described above. We additionally include a relative offset parameter, $\delta_{\rm rel}$, between the NIRISS/SOSS order 1 and 2 spectra. Considering the atmospheric properties, stellar contamination properties, and order 2 vs. 1 offset, our retrieval models have the following number of free parameters: 13 for the atmosphere-only model, 16 for the one-heterogeneity + atmosphere model, and 18 for the star-spots + faculae + atmosphere model. We summarize the priors for each model in Table~\ref{tab:retrieval_summary}. 

\begin{table*}
    \centering
    \caption{Retrieval priors and results from WASP-52\,b's JWST NIRISS/SOSS transmission spectrum.}
    \label{tab:retrieval_summary}
    \renewcommand{\arraystretch}{1.3}
   \begin{tabular}{|l|c|ccc|cc|}
        \hline\hline
        &  & \multicolumn{3}{c|}{\texttt{exoTEDRF}} & \multicolumn{2}{c|}{\texttt{NAMELESS}} \\
        \cmidrule(lr){3-5} \cmidrule(lr){6-7} 
        \textbf{Parameters} & \textbf{Priors} & Atmosphere & Spots & ${\rm Spots + Faculae}$ & Atmosphere & ${\rm Spots + Faculae}$ \\
        \hline
        \textbf{P-T Profile} & & & & & & \\
        ${\rm R_{p, ref}}$ (R$_{\rm Jup}$) & $\mathcal{U}(0.89, 1.46)$ & $1.18^{+0.01}_{-0.01}$ & $1.19^{+0.01}_{-0.02}$ & $1.19^{+0.01}_{-0.01}$ & $1.19^{+0.02}_{-0.01}$ & $1.21^{+0.01}_{-0.01}$ \\
        $\rm{T}$ (K) & $\mathcal{U}(300, 1600)$ & $1075^{+170}_{-161}$ & $876^{+112}_{-117}$ & $1086^{+149}_{-156}$ & $1135^{+188}_{-211}$ & $890^{+208}_{-244}$ \\
        \hline
        \textbf{Composition} & & & & & & \\
        log H$_2$O & $\mathcal{U}(-12, -1)$ & $-3.75^{+0.88}_{-0.57}$ & $-1.34^{+0.22}_{-0.40}$ & $-4.18^{+0.60}_{-0.44}$ & $-4.13^{+0.74}_{-0.41}$ & $-3.90^{+1.41}_{-0.56}$ \\
        log K & $\mathcal{U}(-12, -1)$ & $-8.86^{+1.06}_{-0.87}$ & $-5.16^{+0.87}_{-1.09}$ & $-9.40^{+0.87}_{-0.90}$ & $-9.29^{+0.86}_{-0.87}$ & $-9.12^{+1.69}_{-1.10}$ \\
        log CO$_2$ & $\mathcal{U}(-12, -1)$ & $< -5.49$ & $< -4.47$ & $< -6.16$  & $< -5.55$ & $< -5.30$ \\
        log CH$_4$ & $\mathcal{U}(-12, -1)$ & $< -6.18$ & $< -5.22$ & $< -6.41$ & $< -6.54$ & $< -6.43$ \\
        log NH$_3$ & $\mathcal{U}(-12, -1)$ & $< -5.55$ & $< -4.58$ & $< -5.94$ & $< -6.03$ & $< -6.07$ \\
        log HCN & $\mathcal{U}(-12, -1)$ & $< -4.23$ & $< -2.61$ & $< -4.74$ & $< -3.92$ & $< -2.50$ \\
        \hline
        \textbf{Aerosols} & & & & & & \\
        log $a$ & $\mathcal{U}(-4, 8)$ & $7.37^{+0.44}_{-0.68}$ & $5.61^{+1.78}_{-5.88}$ & $6.65^{+0.82}_{-1.07}$ & $7.43^{+0.41}_{-1.01}$ & $6.06^{+1.26}_{-3.19}$ \\
        $\gamma$ & $\mathcal{U}(-20, 2)$ & $-5.38^{+0.88}_{-0.72}$ & $-4.47^{+1.58}_{-7.88}$ & $-5.81^{+1.43}_{-1.49}$ & $-5.77^{+0.96}_{-0.75}$ & $-6.25^{+2.01}_{-2.31}$ \\
        log P$_{\rm cloud}$ & $\mathcal{U}(-6, 2)$ & $-1.28^{+2.15}_{-2.31}$ & $-3.12^{+3.28}_{-1.81}$ & $-0.80^{+1.79}_{-1.75}$ & $-1.17^{+2.07}_{-2.41}$ & $-1.19^{+2.01}_{-2.01}$ \\
        $\phi_{\rm cloud}$ & $\mathcal{U}(0, 1)$ & $0.56^{+0.06}_{-0.05}$ & $0.60^{+0.08}_{-0.08}$ & $0.49^{+0.08}_{-0.07}$ & $0.49^{+0.06}_{-0.05}$ & $0.42^{+0.13}_{-0.08}$ \\
        \hline
        \textbf{Stellar Heterogeneities} & & & & & & \\
        $f_{\rm spot}$ & $\mathcal{U}(0, 0.5)$ & --- & $0.05^{+0.03}_{-0.02}$ & $0.34^{+0.10}_{-0.12}$ & --- & $0.23^{+0.11}_{-0.07}$ \\
        $f_{\rm fac}$ & $\mathcal{U}(0, 0.5)$ & --- & --- & $0.18^{+0.15}_{-0.08}$ & --- & $0.13^{+0.09}_{-0.05}$ \\
        $T_{\rm spot}$ (K) & $\mathcal{U}(3500, 5500)$ & --- & $3917^{+158}_{-154}$ & $4664^{+94}_{-137}$ & --- & $4479^{+156}_{-266}$ \\
        $T_{\rm fac}$ (K) & $\mathcal{U}(4500, 7500)$ & --- & --- & $5755^{+320}_{-227}$ & --- & $5734^{+271}_{-226}$ \\
        $T_{\rm phot}$ (K) & $\mathcal{N}(5000, 100)$ & --- & $5019^{+87}_{-92}$ & $5111^{+42}_{-35}$ & --- & $5064^{+69}_{-71}$ \\
        \hline
        \textbf{Data offset} & & & & & & \\
        $\delta_{\rm rel}$ & $\mathcal{U}(-500, 500)$ & $201^{+85}_{-84}$ & $301^{+94}_{-89}$ & $148^{+85}_{-82}$ & $177^{+88}_{-87}$ & $111^{+98}_{-92}$ \\
        \hline
        \textbf{Statistics} & 
        ${\chi}^2_\nu$ & 1.03 & 1.07 & 0.91 & 0.85 & 0.74 \\
        & $\ln \mathcal{Z}_{\mathrm{Bayesian}}$ & $1117.4$ & $1114.7$ & $1122.5$ & $1119.1$ & $1123.8$ \\
        & $\mathcal{B}_{01}$ & Ref & $0.07$ & $164$ & Ref & $110$ \\
        & Significance & Ref & N/A & $3.6 \sigma$ & Ref & $3.5 \sigma$ \\
        \hline
    \end{tabular}

\end{table*}

\subsection{Retrieval results} \label{subsec:retrieval_results}

\begin{figure*}
    \centering
    \includegraphics[width=\textwidth]{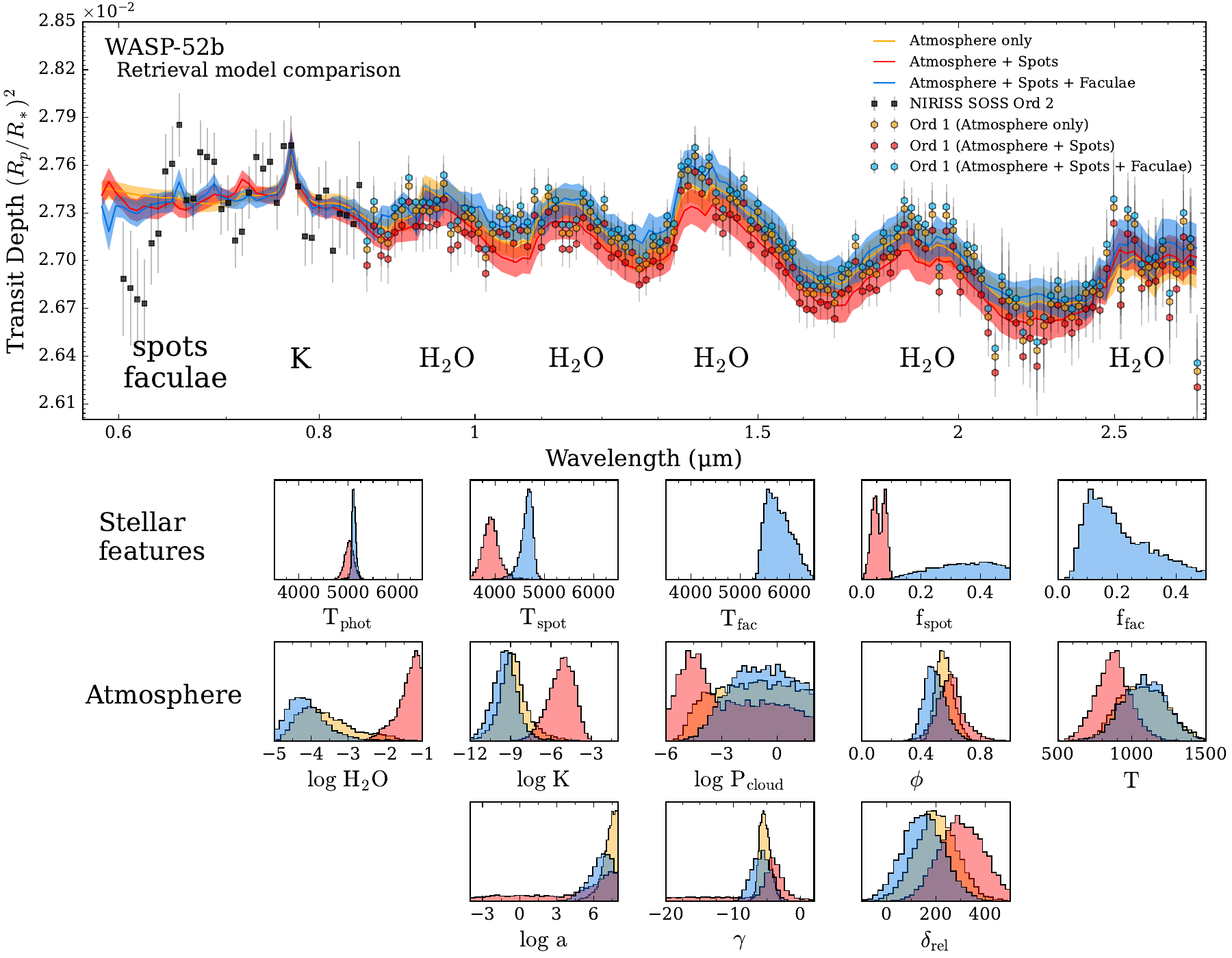}
    \vspace{-0.3cm}
    \caption{Atmospheric and stellar retrieval results for WASP-52\,b. \emph{Top}: NIRISS/SOSS transmission spectrum from the \texttt{exoTEDRF} dataset reduction (order 1: coloured and order 2: black data points), along with the three models overplotted: atmosphere-only (yellow), atmosphere + spots (red), and atmosphere + spots + faculae (blue). The coloured NIRISS/SOSS order 1 data are offset by the best-fitting retrieved offset from each model. The different models are represented by the median retrieved spectrum (solid line) and the $\pm 1\,\sigma$ confidence interval (shaded contours). \emph{Bottom}: Posterior probability distributions corresponding to the three retrieval models. The top row highlights the retrieved stellar contamination parameters, while the middle and bottom rows show the retrieved atmospheric properties (see Table~\ref{tab:retrieval_summary} for the full retrieval results).
    }
    \label{fig:spectra_comparison}
\end{figure*}

WASP-52\,b's JWST NIRISS transmission spectrum can be explained by H$_2$O and K absorption alongside unocculted stellar active regions and atmospheric aerosols. We detect H$_2$O at 10.8\,$\sigma$ confidence and hints of K at 2.5\,$\sigma$ confidence but find no evidence of any other chemical species. Figure~\ref{fig:spectra_comparison} demonstrates that the NIRISS/SOSS data exhibits a spectral slope with increasing transit depth towards short wavelengths, which can be explained by either a scattering haze or unocculted star-spots. When spots (but not faculae) are included in the retrieval model, we find that they partially substitute for the haze as the preferred explanation for the spectral slope in the NIRISS/SOSS data (as discussed further in Section~\ref{subsubsec:results_atmosphere}) and increase the offset between the NIRISS/SOSS orders. However, the model including spots and faculae (preferred at 3.6\,$\sigma$ over the atmosphere-only model and 4.4\,$\sigma$ over the spot-only model) still requires a contribution from atmospheric hazes. Regardless of the inclusion of stellar contamination, we find that an atmospheric haze with an inhomogeneous terminator fraction is needed to explain the observations. We explore these atmospheric and stellar inferences quantitatively in the following subsections.

\subsubsection{Unocculted active regions on WASP-52} \label{subsubsec:results_unocculted}

We first determine whether unocculted active regions are required to explain WASP-52\,b's NIRISS transmission spectrum. Table~\ref{tab:retrieval_summary} (lower rows) summarizes various fit quality and model comparison metrics (reduced chi-squared ${\chi_\nu}^2$, log Bayesian evidence $\ln \mathcal{Z}_{\mathrm{Bayesian}}$, Bayes factors $\mathcal{B}_{01}$ and detection significances) from our retrievals with different stellar contamination model configurations, and data reductions. Across all two transmission spectra (\texttt{exoTEDRF} and \texttt{NAMELESS}), we find that the model, including star-spots and faculae, has the highest Bayesian evidence and the lowest reduced ${\chi}^2$, with equivalent detection significances of $>$ 3.5\,$\sigma$ compared to the atmosphere-only model. However, the atmosphere-only model with no stellar contamination still provides a reasonable fit to the NIRISS/SOSS observations (${\chi_\nu}^2 \approx 1$). Interestingly, there is no improvement in the Bayesian evidence or reduced ${\chi}^2$ when the retrieval configuration switches from atmosphere-only to the atmosphere + star-spots model. This indicates that adding a spectral slope from star-spots provides no additional explaining power compared to an atmospheric haze, so the fitness metrics penalize the additional three parameters introduced by the single stellar spot heterogeneity. The improvement in the fitting metrics from adding faculae arises from the data points with lower transit depths near 0.6\,$\micron$ (see Figure~\ref{fig:spectra_comparison}). Therefore, in what follows, we refer to the atmosphere + star-spots + faculae model as the `preferred' model.

The preferred stellar contamination model indicates the presence of both cold and hot active regions on WASP-52. Our results for \texttt{exoTEDRF} find star-spots and faculae covering $34^{+10}_{-12}$\% and $18^{+15}_{-8}$\% of the visible stellar hemisphere, with corresponding temperatures $\approx$ 450\,K cooler and $\approx$ 650\,K hotter than the stellar photosphere, respectively. We find similar results for \texttt{NAMELESS}, albeit with a lower star-spot covering fraction of $23^{+11}_{-7}$\% (see Table~\ref{tab:retrieval_summary}). These results demonstrate that unocculted active regions are also likely present on WASP-52, alongside the occulted active regions seen in the transit light-curves in Figure~\ref{fig:starspot_lc}, consistent with previous HST observations \citep{bruno2018,bruno2020}. 

\subsubsection{The atmosphere of WASP-52\,b} \label{subsubsec:results_atmosphere}

\begin{figure*}
    \centering
    \includegraphics[width=\textwidth]{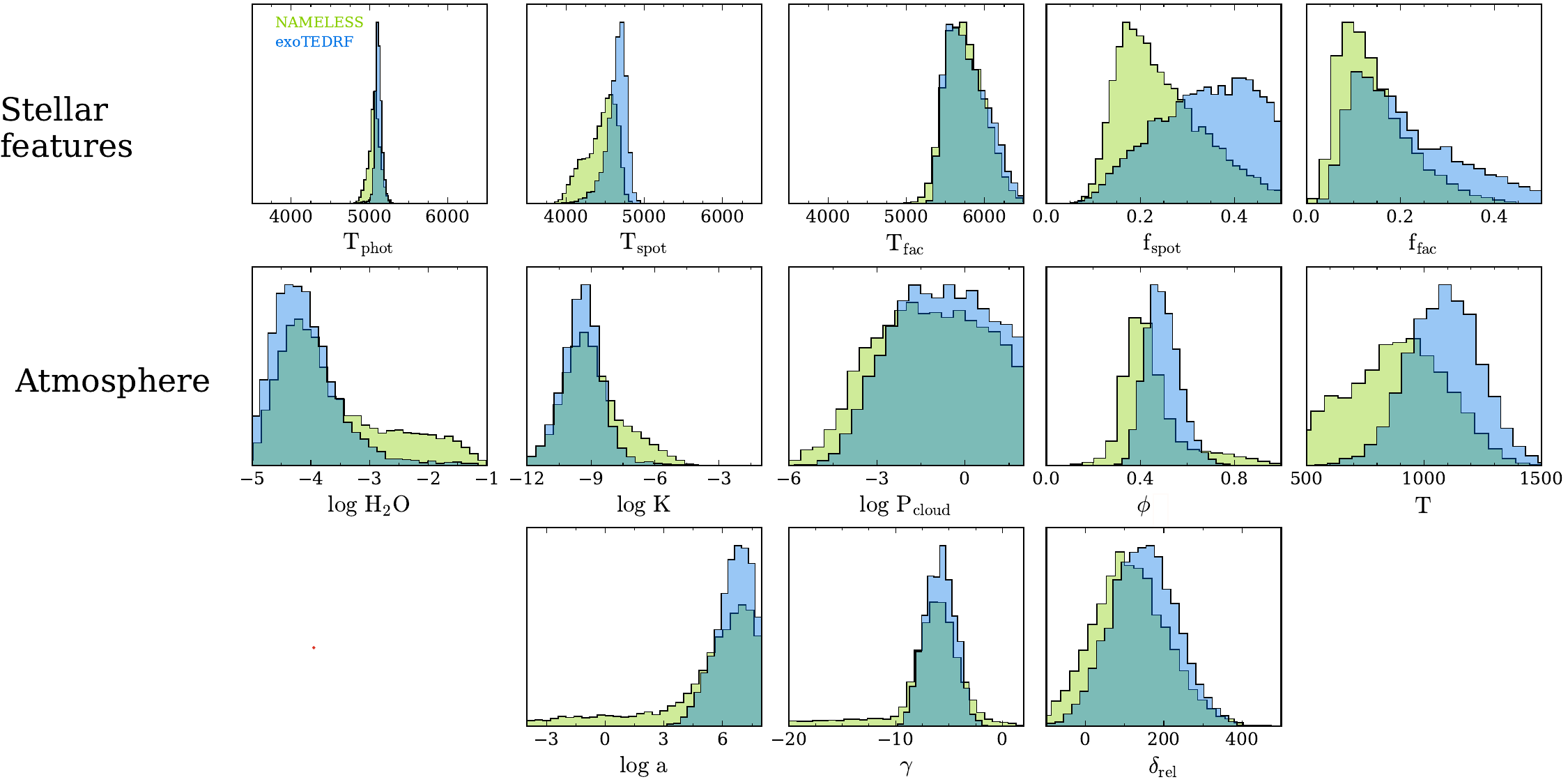}
    \vspace{-0.3cm}
    \caption{Sensitivity of retrieval results to dataset reduction. The posterior distributions for \texttt{exoTEDRF} (blue) and \texttt{NAMELESS} (green) are shown for the statistically preferred atmosphere + spots + faculae model. \emph{Top}: Retrieved stellar contamination properties. \emph{Middle and bottom}: Key retrieved atmospheric properties (see Table~\ref{tab:retrieval_summary} for the full retrieval results).
    }
    \label{fig:dataset_comparison}
\end{figure*}

We next report our atmospheric retrieval inferences and their sensitivity to stellar contamination model choices and data reductions (\texttt{exoTEDRF} vs. \texttt{NAMELESS} pipeline). Figure~\ref{fig:spectra_comparison} presents our retrieval results for the \texttt{exoTEDRF} for the three stellar contamination models, while Figure~\ref{fig:dataset_comparison} shows the sensitivity of our retrieval results to the two transmission spectra for the statistically preferred retrieval model (star-spots + faculae). We also provide the full retrieval results across all model and data combinations in Table~\ref{tab:retrieval_summary}.

Our preferred atmospheric retrieval model favours sub-solar-to-solar H$_2$O and K abundances. We find broadly consistent H$_2$O abundances for both the atmosphere-only (\texttt{exoTEDRF}, atmosphere-only: $\log \mathrm{H_2 O} = -3.75^{+0.88}_{-0.57}$) and atmosphere + star-spots + faculae models (\texttt{exoTEDRF}, star-spots + faculae: $\log \mathrm{H_2 O} = -4.18^{+0.60}_{-0.44}$), which are consistent with either a solar ($\log \mathrm{H_2 O}_{\rm{solar}} \approx -3.3$; \citealt{Asplund2021}) or somewhat subsolar atmospheric metallicity. For the preferred retrieval model, we also find excellent agreement between \texttt{exoTEDRF} (star-spots + faculae: $\log \mathrm{H_2 O} = -4.18^{+0.60}_{-0.44}$) and \texttt{NAMELESS}  (star-spots + faculae: $\log \mathrm{H_2 O} = -3.90^{+1.41}_{-0.56}$). Similarly, we find sub-solar K abundances (e.g., $\log \mathrm{K} = -9.40^{+0.87}_{-0.90}$ for \texttt{exoTEDRF} under the star-spots + faculae model, vs. $\log \mathrm{K}_{\rm{solar}} \approx -6.9$; \citealt{Asplund2021}). While we do not detect other gases, our retrievals place strong 2\,$\sigma$ upper limits on the abundances of CO$_2$, CH$_4$, and NH$_3$ ruling out abundances exceeding 10\,ppm (see Table~\ref{tab:retrieval_summary}).

Atmospheric hazes also play a role in shaping WASP-52\,b's NIRISS transmission spectrum. Our retrievals consistently favour a strongly scattering haze $\sim\,10^6$ stronger than H$_2$ Rayleigh scattering with a scattering power law exponent of $\sim\,-6$ (see Table~\ref{tab:retrieval_summary}). The hazes are distributed inhomogeneously around WASP-52\,b's terminator with a covering fraction of $\approx 50\,\%$ (e.g., $\phi = 0.49^{+0.08}_{-0.07}$ for \texttt{exoTEDRF} under the star-spots + faculae model). We do not find evidence for an optically thick high-altitude cloud deck for our preferred retrieval model with both star-spots and faculae.

Under the assumption of only unocculted star-spots (with no faculae), we find an unphysical atmospheric solution with a significantly higher H$_2$O abundance than would be expected for a hot Jupiter like WASP-52\,b ($\log \mathrm{H_2 O} \sim -1$, or 10\% H$_2$O; see Figure~\ref{fig:spectra_comparison}). Besides the high H$_2$O abundance, this solution requires a lower atmospheric temperature ($\approx$ 900\,K vs. $\approx$ 1100\,K for the star-spots + faculae model), a large $301^{+94}_{-89}$\,ppm offset between the NIRISS orders, and star-spots over 1000\,K cooler than the stellar photosphere covering $5^{+3}_{-2}$\% of the stellar disk. We regard this extreme solution as unlikely --- arising from the complex degeneracy between atmospheric hazes, unocculted star-spots, and the relative offset between the NIRISS orders --- with the lower retrieved offsets for the preferred star-spots + faculae model ($148^{+85}_{-82}$\,ppm for \texttt{exoTEDRF} and $111^{+98}_{-92}$\,ppm for \texttt{NAMELESS}) as additional lines of evidence that both star-spots and faculae must be considered simultaneously for reliable atmospheric inferences.

\section{Detection of excess helium absorption}  \label{sec:He}

We search for evidence of upper-atmosphere helium absorption by independently analyzing the pixel resolution light-curves and transmission spectrum around 1.083\,$\micron$. Due to the escaping nature of this atmospheric tracer leading to potential pre- and post-transit absorption, it is mandatory to analyze the line using a data-driven approach.
 
\begin{figure}
    \centering
    \includegraphics[width=\columnwidth]{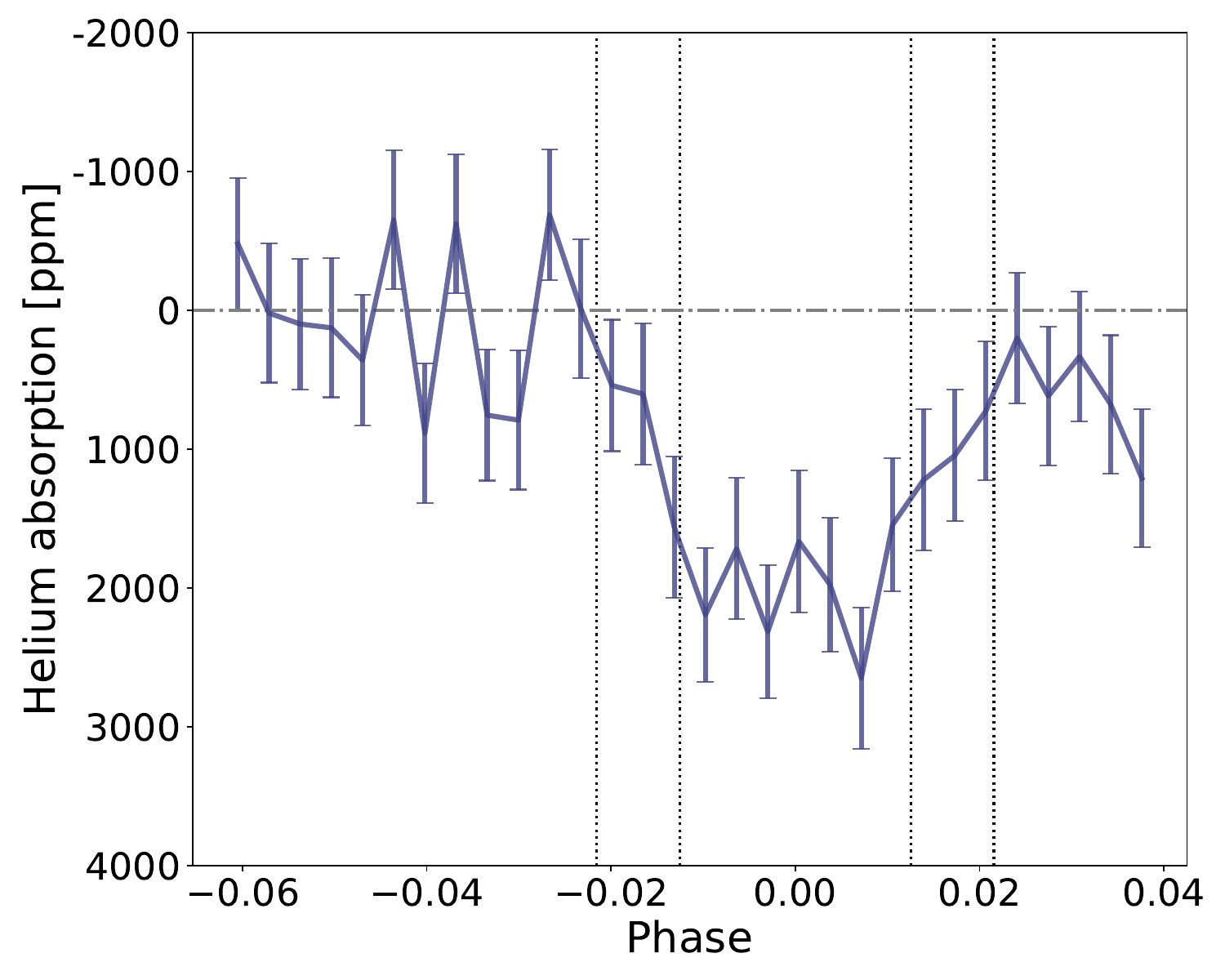}
    \vspace{-0.3cm}
    \caption{Relative helium absorption light-curve. The transit contact points t$_1$, t$_2$, t$_3$ and t$_4$ are shown as vertical dotted lines from left to right. The grey horizontal dashed line indicates the lack of excess absorption compared to the reference light-curve.}
    \label{fig:helium_LC}
\end{figure}

We first analyze the light-curves at the pixel resolution of NIRISS/SOSS for the 47 pixels around the helium pixel (centered at 1.083\,$\micron$), covering the 1.062--1.105\,$\micron$ wavelength range. We build a reference light-curve surrounding the helium triplet by averaging all the light-curves (a total of 38) from 1.062 to 1.078\,$\micron$ and from 1.088 to 1.105\,$\micron$. Therefore, this reference light-curve is not biased by escaping helium. A linear trend is visible in the out-of-transit baseline of the 47 individual and reference light-curves. We thus fit this trend using the reference light-curve and subtract it from all the light-curves assuming that no significant variation of the amplitude of the linear trend occurs over this narrow wavelength range. Then, each light-curve is normalized by its average out-of-transit flux measured on exposures taken at phase <\,-0.04, which reduces the impact of any extended helium atmosphere beyond transit on the final helium light-curve. We then compute relative light-curves by dividing them by the detrended reference light-curve to extract the excess absorption at each pixel; the result is shown in Figure \ref{fig:helium_LC}. We see clear absorption of $\sim$\,2000\,ppm during the transit, and additional post-transit absorption evidenced at $\sim$\,2.9\,$\sigma$ (measured as the average flux post transit), indicating potential atmospheric escape beyond the Roche lobe in the form of a cometary-like tail. However, the lack of longer post-transit absorption makes it difficult to robustly confirm the presence of a cometary-like tail and estimate its duration. As a verification, we inspected the surrounding relative light-curves of the helium pixel, but they did not show any detectable absorption. 

\begin{figure}
    \centering
    \includegraphics[width=\columnwidth]{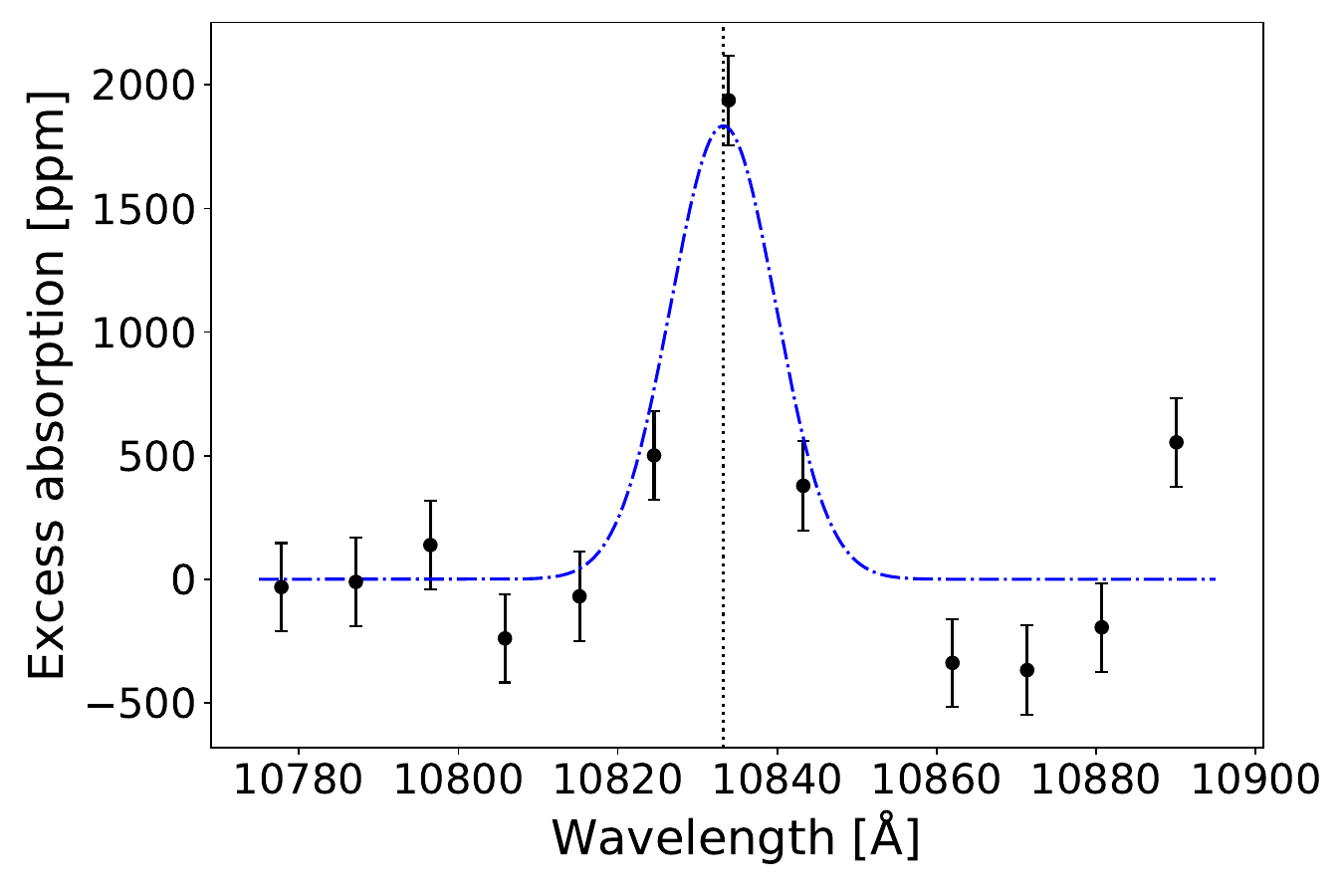}
    \vspace{-0.3cm}
    \caption{Transmission spectrum around the helium triplet (indicated with the vertical dotted line), alongside the best Gaussian fit with a width fixed to 0.75\,\AA\ once convolved at the native resolution (blue).}
    \label{fig:helium_TS}
\end{figure}

In a second step, we build the transmission spectrum by measuring the absorption between transit contact points t$_2$ and t$_3$ for the 47 relative light-curves, which is shown in Figure \ref{fig:helium_TS}. This is preferred to the pixel resolution transmission spectrum derived in Section\,\ref{sec:data}, as the latter is biased by the presence of the post-transit absorption signal. We confirm a clear signal of 1916 $\pm$ 264\,ppm (7.3\,$\sigma$) by fitting a Gaussian of free amplitude and full width at half maximum (FWHM). Based on the work done in \citet{fournier-tondreau2024}, \citet{radica2024} and \citet{Piaulet_Ghorayeb_2024}, we also consider the following model: a Gaussian with a fixed width (0.75\,\AA) convolved at the native resolution of NIRISS/SOSS ($R$ = 700), and at the expected location of the He I triplet absorption (10833.33\,\AA). This model provides an estimate of the resolved helium signature at high spectral resolution. The best-fit value for the helium amplitude expected at high-resolution is 5.5 $\pm$ 0.9\,\%. This is easily within reach of current high-resolution spectrographs for more robust detection and interpretation of the helium line shape, further constraining the mass-loss rate and dynamics. As shown in \cite{fu2023}, modelling of the unresolved helium triplet with JWST leads to significant degeneracy between the mass-loss rate and the thermospheric temperature. Thus, the optimal way to proceed to achieve a complete understanding of WASP-52\,b upper atmosphere is through a combined modelling of JWST and a high-resolution spectrograph to have the signal fully resolved temporally and spectrally, out of the scope of this paper.

\citet{kirk2022} were the first to report the detection of helium in WASP-52\,b's atmosphere at high-resolution, whereas \citet{vissapragada2020} had reported an upper limit at low resolution. \citet{kirk2022} reported an excess absorption of 3.44 $\pm$ 0.31\,\% with the NIRSPEC instrument on the Keck II telescope and derived a mass-loss rate of $\sim$\,1.4 $\times$ 10$^{11}$\,g\,s$^{-1}$ using the \texttt{p-winds} code \citep{dossanto2022}. More recently, \citet{allart2023} with the high-resolution spectropolarimeter SPIRou on the Canada–France–Hawaii Telescope (CFHT) found an upper limit at 1.69\,\% that disagreed with the previously published results. Our detection agrees with \cite{kirk2022}, which contrasts with \cite{allart2023}. Given that WASP-52 is a young active star with intense XUV irradiation \citep{allart2023}, it might be possible that the helium signature of WASP-52\,b is variable in time under the change of its stellar environment. Further high-resolution and JWST observations are thus required to fully characterize the escaping atmosphere of WASP-52\,b.

\section{Summary \& discussion}\label{sec:discu}

We presented here the first JWST transmission spectrum of WASP-52\,b, a hot Saturn-mass exoplanet orbiting an active K dwarf. Our NIRISS/SOSS observations required a careful analysis of stellar active regions, both in terms of crossing events during light-curve fitting and contamination of the 0.6--2.8\,$\micron$ transmission spectrum during atmospheric retrieval. Our main results are as follows:
    
\begin{itemize}
    \item We detect two spot-crossing events with NIRISS/SOSS. We thus extracted the transmission spectrum using a spot-transit model to infer the properties of the occulted active regions on the host star WASP-52. The two occulted star-spots, one before the mid-transit and another after, cover together about 2.4\,\% of the stellar surface with temperatures $\approx$ 400--500\,K colder than the photosphere. The best-fitting parameters (position, size, temperature) of the second spot are well constrained, but the $y$-position of the first spot, orthogonal to the path of the planet's transit, is degenerate and correlated with its size and temperature. We found setting the star-spot surface gravity as a free parameter was unnecessary.
    \item Our retrieval analysis finds that a model including unocculted spots and faculae, jointly with a planetary atmosphere, provides the best fit to WASP-52\,b's transmission spectrum. Spots and faculae are detected at 3.6\,$\sigma$ compared to an atmosphere-only model. The unocculted star-spots and faculae cover $\approx\,30 \pm 10\,$\% and $\approx\,20 \pm 10\,$\% of the visible stellar hemisphere, respectively, with corresponding temperatures $\approx$ 450\,K cooler and $\approx$ 650\,K hotter than the stellar photosphere, respectively. The retrieved temperature of the unocculted star-spots is consistent with the inferred temperature of the occulted star-spots.
    \item We detect multiple strong atmospheric absorption features caused by H$_2$O vapor (10.8\,$\sigma$). The retrieved H$_2$O abundances suggest a sub-solar or solar atmospheric metallicity ($\log$ H$_2$O = -4.18$^{+0.60}_{-0.44}$ for the \texttt{exoTEDRF} reduction; $\log$ H$_2$O = -3.90$^{+1.41}_{-0.56}$ for the \texttt{NAMELESS} reduction).
    \item We additionally find evidence of K absorption (2.5\,$\sigma$) and a slope towards shorter wavelengths consistent with scattering from atmospheric hazes.
    \item Finally, we detect atmospheric He with an excess absorption near 1.083\,$\micron$ of 1916 $\pm$ 264\,ppm. A tentative signature of additional post-transit absorption is also present in the light-curve surrounding the He triplet ($\sim$2.9\,$\sigma$), hinting at potential atmospheric escape.
\end{itemize}

We proceed to discuss the broader context surrounding our findings and the implications of our results.

\subsection{The atmosphere of WASP-52 b in context}

Our results confirm the presence of H$_2$O vapour reported in previous works using the Hubble Space Telescope \citep{tsiaras2018,bruno2018,bruno2020}, alongside K and He from ground-based observations \citep{chen2020,kirk2022}. Our best-fitting retrieval model (including star-spots and faculae) finds a H$_2$O abundance of $\log$ H$_2$O = -4.18$^{+0.60}_{-0.44}$ (\texttt{exoTEDRF}) or $\log$ H$_2$O = -3.90$^{+1.41}_{-0.56}$ (\texttt{NAMELESS}), which is consistent within 1\,$\sigma$ to the H$_2$O abundances reported from previous Hubble observations ($\log$ H$_2$O = $-3.30^{+0.94}_{-1.12}$ \citep{bruno2020} and log H$_2$O = -4.09 $\pm$ 0.87 \citep{tsiaras2018} --- though the latter study did not account for stellar activity). Our most precise H$_2$O abundance (from the \texttt{exoTEDRF} reduction) is subsolar to 2\,$\sigma$ (compared to a solar value of $\log$ H$_2$O = -3.3; \citealt{Asplund2021}), which is suggestive of either a sub-solar atmospheric metallicity or a super-solar C/O ratio. We note that the wider uncertainties on the H$_2$O abundance from the \texttt{NAMELESS} reduction allow a solar H$_2$O abundance. Ultimately, additional observations at longer infrared wavelengths are necessary to detect other chemical tracers, such as CO$_2$ and CO, to complete the chemical inventory of WASP-52\,b’s atmosphere and allow the study of more complex effects (such as disequilibrium chemistry and vertical mixing) and constrain the planet's formation history.

Our inference of a subsolar or solar H$_2$O abundance for WASP-52\,b is consistent with the mass-metallicity trend seen for other hot giant exoplanets. In the solar system, a clear inverse trend is observed between planetary mass and atmospheric metallicity (using CH$_4$ as a proxy; \citealt{Atreya2018}), which is interpreted as evidence of formation via core accretion \citep{Pollack1996}. With a mass similar to Saturn and a host star with solar metallicity ([Fe/H] = 0.03 $\rm{\pm}$ 0.12; \citealt{hebrard2013}), WASP-52\,b would be expected to have an atmospheric metallicity near the value of 10\,$\times$ solar inferred from Saturn's CH$_4$ abundance \citep{Fletcher2009,Atreya2018}. However, population analyses of hot giant exoplanets have generally favoured a lower trendline for H$_2$O abundances compared to CH$_4$ abundances in the solar system \citep[e.g.,][]{Welbanks2019,Sun2024}. The most recent exoplanet mass-metallicity trend presented by \citet{Sun2024} predicts an atmospheric O abundance for a 0.46\,$M_{\rm J}$ planet almost identical to the stellar O abundance, in good agreement with our subsolar to solar H$_2$O abundance.

Atmospheric hazes are needed to explain WASP-52\,b's JWST NIRISS transmission spectrum. We also find that a haze is required regardless of the inclusion of unocculted star-spots, which can mimic the signature of a scattering haze. Observations of muted spectral features in WASP-52\,b's transmission spectra were previously attributed to an optically thick cloud deck \citep{kirk2016,chen2017,alam2018}, but the retrieval analysis on the combined HST/STIS, WFC3, and \textit{Spitzer}/IRAC transmission spectrum performed by \citet{bruno2020} was compatible with solutions with and without clouds. Our JWST retrieval analysis finds no evidence of optically thick high-altitude gray clouds, suggesting the primary aerosols in WASP-52\,b's upper atmosphere take the form of small particle scattering hazes. Ultimately, our findings of haze scattering in WASP-52\,b's atmosphere are compatible with theoretical work on aerosols and condensates, along with observational trends that suggest that hot Jupiters near WASP-52\,b's equilibrium temperature should have hazy atmospheres \citep[e.g.,][]{barstow2017}.

Additional observations will help us better understand the atmospheric chemical compositions and aerosol properties of WASP-52\,b. The JWST GO 3969 Program has recently obtained a NIRSPec/G395H transit spectrum for this planet ranging from 2.8-5.2\,$\mu$m. Combining the NIRSpec and NIRISS observations, will allow detailed constraints on the atmospheric inventory on key O, C, and N-bearing molecules, including $\rm{HCN}$ near 3.1\,$\micron$ \citep{macDonald2017HCN}, $\rm{H_2 S}$ near 4.0\,$\micron$ \citep{fu2024}, $\rm{CO_2}$ near 4.3\,$\micron$ \citep{ers2023}, and $\rm{CO}$ near 4.7\,$\micron$ \citep{rustamkulov2023}. 

\subsection{Accounting for the stellar activity of WASP-52} 

\subsubsection{Occulted star-spots impacting transit light-curves}

In this work, we jointly retrieved the parameters of the occulted active regions and the planetary transit. Two independent light-curve fitting methods effectively constrained the properties of the spot that was occulted after mid-transit.  However, similarly to the findings in \citet{fournier-tondreau2024}, our results for the first occulted spot indicate lingering uncertainties regarding the inferred star-spot properties. Deducing the two-dimensional distribution of heterogeneities from transit light-curves is a degenerate problem since varying the spot $y$-positions (orthogonal to the transit chord) can yield identical light-curves. Considering the correlation between the position of the spot and its size and temperature, this can lead to different spot properties. Additionally, the different assumptions within the spot-transit model may hinder some solutions from being ruled out. Specifically, \texttt{spotrod} assumes circular, homogeneous active regions (without umbra and penumbra structures for spots, nor elongated shapes for faculae), which follow the star limb darkening law \citep{beky2014}. Moreover, we explored whether using all spectral channels to fit the spot position and radius would lift some degeneracies. Nevertheless, this method did not provide a unique solution for the first spot, which was also loosely constrained with the standard light-curve fitting method. Fortunately, regardless of the occulted spot correction method, our retrieval results between the derived \texttt{exoTEDRF} and \texttt{NAMELESS} transmission spectra are in good agreement despite using different spot parameters (see Table \ref{tab:wlc_parameters}). 

Our work highlights the benefits of modelling star-spots in transit light-curves. First, the NIRISS/SOSS light-curves were significantly deformed by the spot-crossings and therefore masking them would decrease the observing efficiency. Second, the temperatures inferred for the occulted spots agree with those independently derived from our retrieval modelling of unocculted spots, which provides reassurance that both methods are accurately inferring properties of WASP-52's active regions. Furthermore, the retrieved covering fraction of occulted spots is expected to be much lower than that of unocculted spots, as the proportion of the photosphere swept by the planet's transit chord is relatively small. Observational data suggest that active K-type dwarfs have an overall spot coverage between 20-35\,\% \citep[e.g.,][]{Nichols-Fleming2020}, which is consistent with our results ($f_{\rm spot} \approx 30 \pm 10$\,\% for \texttt{exoTEDRF}).

\subsubsection{Unocculted active regions contaminating transmission spectra}

The significant information gained from JWST NIRISS compared to its predecessor enables us to mitigate the size-temperature degeneracy of unocculted active regions, a limitation faced in earlier observations with Hubble \citep{bruno2020}. This could explain why the inferred unocculted spot properties for our preferred model ($\Delta T_{\rm spot} \approx$ 450\,K with $f_{\rm spot} \approx 30 \pm 10$\,\% for \texttt{exoTEDRF}) are not in agreement with those reported in \citet{bruno2020}, which found significantly cooler spots with a lower coverage fraction ($T_{\rm spot}$ < 3000\,K with $f_{\rm spot}$ = 5\,\%). Recently, \citet{savanov2019} showed, using three independent observational sources, that spot temperatures for active stars with $T_{\rm{eff}} \sim$\,5000\,K are expected to be about 750\,K cooler than the stellar photosphere, which is in good agreement with our results, but contrasts the earlier trend reported by \citet{berdyugina2005}. Additionally, \citet{bruno2020} only accounted for unocculted spots, while our analysis indicates that the preferred retrieval model favours both unocculted spots and faculae rather than only spots. This adds to the growing body of evidence that employing a single population of unocculted active regions may be an overly simplified prescription of surface heterogeneities \citep[e.g.,][]{zhang2018,lim_atmospheric_2023,fournier-tondreau2024}. 

In this work, we found that the treatment choice for unocculted active regions can strongly impact the retrieved atmospheric properties. Specifically, adopting a model with only a single stellar heterogeneity biased our retrieved H$_2$O abundances to unphysically high values via a complex degeneracy between unocculted star-spots, atmospheric hazes, and the atmospheric mean molecular weight. We also did not find any benefit in setting the surface gravities of active regions as a free parameter for WASP-52, unlike the findings of \citet{fournier-tondreau2024} for HAT-P-18\,b. However, we stress that these joint stellar and planetary retrievals rely on grids of 1D stellar models, which in turn limits their accuracy to the precision of these models \citep[e.g.,][]{iyer2020}.

\subsubsection{Implications for stellar contamination modelling in exoplanet atmospheric retrievals} 

Stellar contamination has rapidly become a key consideration in atmospheric retrievals with the advent of JWST. In particular, searches for rocky planet atmospheres have been hindered by the challenges of stellar contamination \citep{lim_atmospheric_2023,moran2023,radica2024b}. WASP-52\,b provides an ideal case study to assess methods for disentangling stellar contamination from atmospheric features, as both effects have similar magnitudes in WASP-52\,b's transmission spectrum. Our results demonstrate that multiple stellar contamination models, including independent populations of cold and hot active regions, should be considered in atmospheric retrievals.

\section{Conclusion}\label{sec:conclu}
The hot gas giant WASP-52\,b stands out as a key benchmark for atmospheric characterization and stellar contamination mitigation in the JWST era. In this work, we present its first JWST observation while addressing the challenges posed by the host star’s activity in characterizing the exoplanet's atmosphere. Our NIRISS/SOSS observation supports previous water, helium and potassium detections in WASP-52\,b's atmosphere and a subsolar or solar atmospheric metallicity. We identify both occulted star-spots and unocculted active regions (both spots and faculae) and emphasize the necessity of accounting for stellar activity when performing atmospheric characterization through transmission spectroscopy of exoplanets orbiting active stars. Specifically, our results highlight the importance of exploring different stellar contamination models to ensure accurate atmospheric inferences and avoid potential biases. 

Additionally, our study points to promising future avenues. First, the upcoming analysis of the JWST NIRSpec observation will expand the chemical inventory of WASP-52\,b and refine the constraints on its atmospheric composition and aerosol properties. Second, the degeneracies between stellar contamination and atmospheric parameters call for the continued development of robust models and strategies to better mitigate stellar contamination in exoplanet observations.

\section*{Acknowledgements}
This work is based on observations made with the James Webb Space Telescope. This project is undertaken with the financial support of the Canadian Space Agency. This research was enabled in part by support provided by Calcul Québec (\url{www.calculquebec.ca}) and the Digital Research Alliance of Canada (\url{alliancecan.ca}). This work benefited from financial support from the Fonds de Recherche du Québec — Nature et technologies (FRQNT) and Natural Sciences and Engineering Research Council (NSERC). M.F.T. acknowledges financial support from the Clarendon Fund Scholarship and the FRQNT. K.M. acknowledges financial support from the FRQNT. R.A. acknowledges the Swiss National Science Foundation (SNSF) support under the Post-Doc Mobility grant P500PT\_222212 and the support of the Institut Trottier for Research on Exoplanets (iREx). R.J.M acknowledges support from NASA through the NASA Hubble Fellowship grant HST-HF2-51513.001, awarded by the Space Telescope Science Institute, which is operated by the Association of Universities for Research in Astronomy, Inc., for NASA, under contract NAS 5-26555. C.P.-G. acknowledges support from the E. Margaret Burbidge Prize Postdoctoral Fellowship from the Brinson Foundation. S.P. acknowledges the financial support of the Swiss National Science Foundation (within the framework of the National Centre of Competence in Research PlanetS supported under grant 51NF40\_205606). L.D. acknowledges support from the NSERC and the Trottier Family Foundation. N.B.C. acknowledges support from an NSERC Discovery Grant, a Tier 2 Canada Research Chair, and an Arthur B. McDonald Fellowship. N.B.C. also thank the Trottier Space Institute and the iREX for their financial support and dynamic intellectual environment. D.J. is supported by the National Research Council (NRC) Canada and by an NSERC Discovery Grant.  

\section*{Data availability}
All data used in this study are publicly available from the Barbara A. Mikulski Archive for Space Telescopes\footnote{\url{https://mast.stsci.edu/portal/Mashup/Clients/Mast/Portal.html}}.



\bibliographystyle{mnras}
\bibliography{W52b.bib} 



\appendix
\section{Additional materials}
\begin{figure*}
	\centering
	\includegraphics[width=\textwidth]{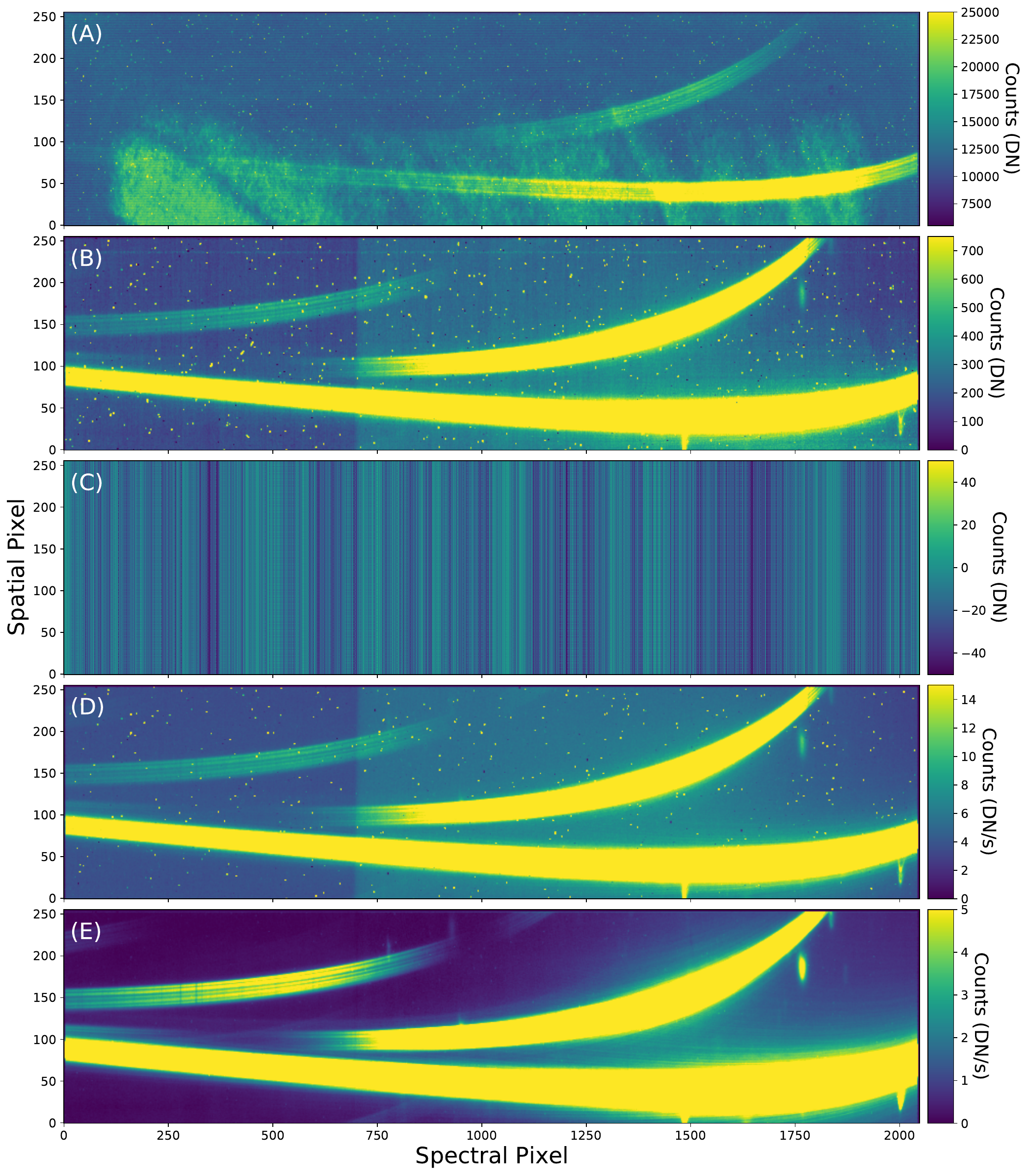}
    \caption{Data products at different stages of the reduction process.
    \textbf{(A)}: Raw, uncalibrated data frame of the 8$^\mathrm{th}$ group of the 20$^\mathrm{th}$ integration in data numbers (DN).
    \textbf{(B)}: Same data frame as (A) after superbias subtraction. 
    \textbf{(C)}: 1/$f$ noise in frame B.
    \textbf{(D)}: Intregration 20 after ramp fitting and flat field correction. 
    \textbf{(E)}: Final calibrated median stack of all out-of-transit integrations.
    \label{fig:Reduction Steps}}
\end{figure*}

\begin{figure*}
	\centering
	\includegraphics[width=\textwidth]{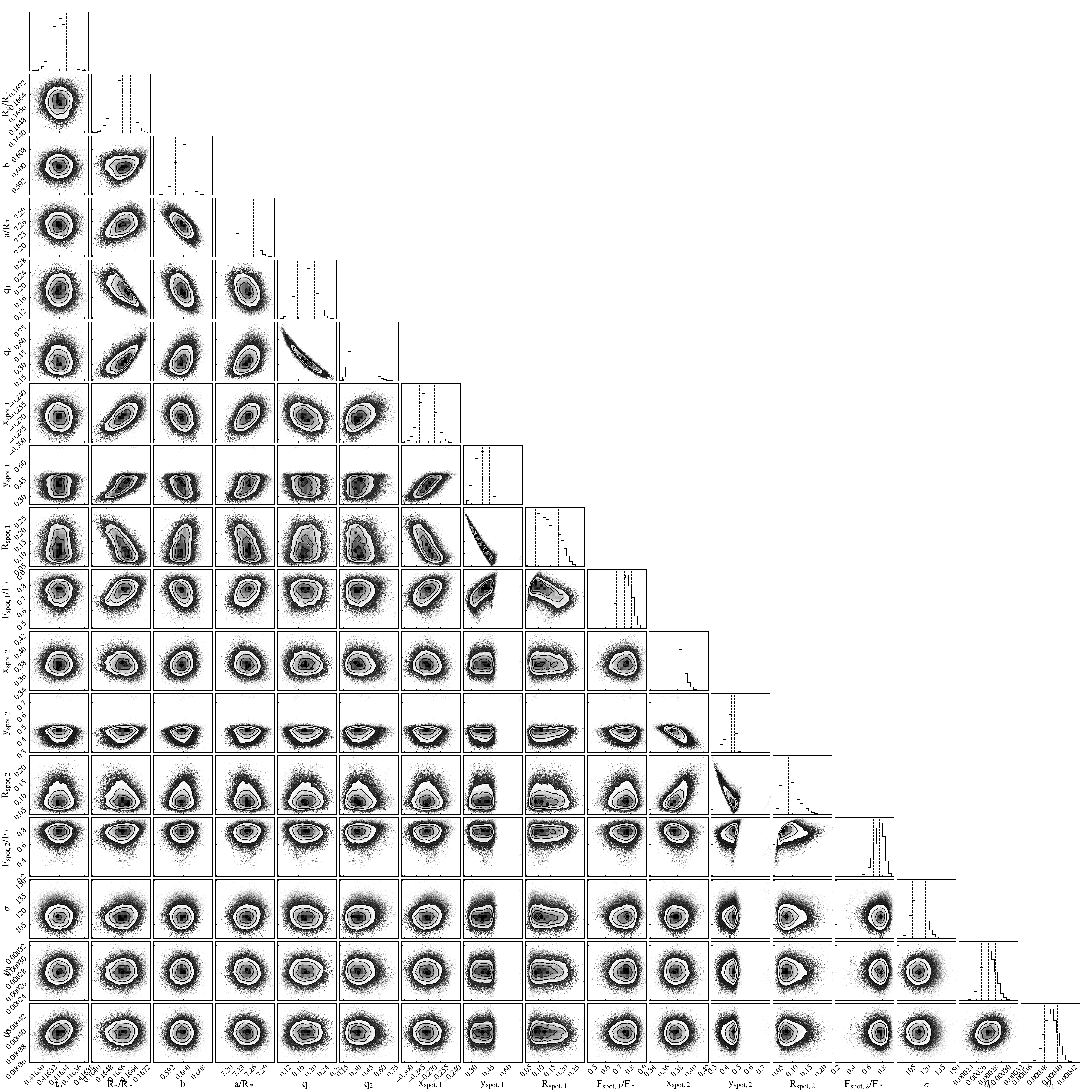}
    \caption{Posterior probability distributions from the broadband light-curve fit of the \texttt{exoTEDRF} dataset reduction. Note that the units of $t_0$ are displayed as BJD -  2459910.
    \label{fig:corner_wlc}}
\end{figure*}

We provide supplementary materials related to the data reduction process (data products shown in Figure \ref{fig:Reduction Steps}) and the fitting of the broadband light-curve (priors and results detailed in Table \ref{tab:wlc_parameters}, and corner plot in Figure \ref{fig:corner_wlc}).

\begin{table*}
    \centering
    \caption{Broadband light-curve fit priors and results from WASP-52\,b's JWST NIRISS/SOSS transit light-curves.}
    \label{tab:wlc_parameters}
    \renewcommand{\arraystretch}{1.5}
    \begin{tabular}{|l|c|cc|cc|} 
    \hline
    \hline
     &&  \multicolumn{2}{c}{\texttt{exoTEDRF}} & \multicolumn{2}{c}{\texttt{NAMELESS}}  \\
     \cmidrule(lr){3-4} \cmidrule(lr){5-6} 
    \textbf{Parameters} & \textbf{Priors} & Highest Likelihood&Median&Highest Likelihood&Median\\
    \hline 
    \textbf{Transit} & &&&& \\
    $\rm t_0$ ( BJD - 2400000) & $\mathcal{U}$\,[59910.39, 59910.44] &59910.416339& 59910.416339	 	 
    $^{+0.000011}_{-0.000011}$  &59910.416346& 59910.416341
    $^{+0.000014}_{- 0.000014}$ \\\\	
    $\rm R_p/R_*$ & $\mathcal{U}$\,[0.01, 0.9] &0.1658&               0.1659$^{+0.0005}_{-0.0005}$ &0.1662&0.1665$^{+0.0004}_{-0.0004}$ \\\\
    $\rm b$ & $\mathcal{U}$\,[0.01, 0.9] &0.598  & 
    0.599$^{+0.003}_{-0.003}$ &0.598& 0.600$^{+0.004}_{-0.004}$  \\\\
    $\rm a/R_*$ & $\mathcal{U}$\,[1, 20]&7.246& 7.250$^{+0.018}_{-0.017}$ &7.26& 7.25$^{+0.02}_{-0.02}$  \\\\ 
    $\rm q_1$ & $\mathcal{U}$\,[0, 1] &0.20 & 0.18$^{+0.03}_{-0.03}$ &0.18& 0.16$^{+0.02}_{-0.02}$ \\\\
    $\rm q_2$ & $\mathcal{U}$\,[0, 1] &0.29& 0.34$^{+0.11}_{-0.09}$ &0.31&0.42$^{+0.10}_{-0.08}$ \\\\ 
    $\rm \theta_0$ & $\mathcal{U}$\,[-10, 10] &0.000283& 0.000279$^{+ 0.000011}_{-0.000011}$ &0.000287&0.000268$^{+0.000014}_{-0.000014}$ \\\\
    $\rm \theta_1$ & $\mathcal{U}$\,[-10, 10] &0.000409 &0.000398$^{+ 0.000009}_{-0.000009}$ &0.000400&0.000391$^{+0.000011}_{-0.000011}$ \\\\  	
    $\rm \sigma$ (ppm) & $\mathcal{L}$\,[0.1, 100000] &107 & 113$^{+7}_{-6}$  &169& 182$^{+8}_{-7}$ \\\\\hline 
    \textbf{Occulted star-spots} & &&&& \\
    $\rm x_{spot,1}$ ($\rm R_*$)& $\mathcal{U}$\,[-1, 0]  &-0.273& -0.272$^{+0.010}_{-0.009}$&-0.259& -0.258$^{+0.010}_{-0.009}$ \\\\
    $\rm y_{spot,1}$ ($\rm R_*$)& $\mathcal{U}$\,[0, 1] &0.38&0.39$^{+0.06}_{-0.07}$&0.48&0.48$^{+0.2}_{-0.03}$\\\\
    $\rm R_{spot,1}$ ($\rm R_*$)& $\mathcal{U}$\,[0, 1] &0.14 &0.13$^{+0.05}_{-0.04}$&0.061& 0.078$^{+0.019}_{-0.017}$\\\\
    $\rm F_{spot,1}/F_*$ & $\mathcal{U}$\,[0, 1] &0.72 &0.75$^{+0.05}_{-0.06}$&0.72& 0.76$^{+0.07}_{-0.11}$  \\\\ 
    $\rm x_{spot,2}$ ($\rm R_*$) & $\mathcal{U}$\,[0, 1]  & 0.370 & 0.377$^{+0.010}_{-0.009}$&0.389&0.390$^{+0.011}_{-0.012}$\\\\ 
    $\rm y_{spot,2}$ ($\rm R_*$)& $\mathcal{U}$\,[0, 1]  &0.47&0.45$^{+0.03}_{-0.04}$&0.41&0.41$^{+0.05}_{-0.05}$\\\\  $\rm R_{spot,2}$ ($\rm R_*$)& $\mathcal{U}$\,[0, 1] 
    & 0.066&0.085$^{+0.03}_{-0.019}$&0.12&0.12$^{+0.04}_{-0.03}$\\\\
    $\rm F_{spot,2}/F_*$ & $\mathcal{U}$\,[0, 1] 
    &0.71&0.76$^{+0.06}_{-0.07}$&0.75& 0.72$^{+0.08}_{-0.09}$ \\\\\hline 
    \multicolumn{6}{l}{\footnotesize \emph{Note:} The set of parameters with the highest likelihood comes from the weighted samples.}\\
    \end{tabular}
\end{table*}

\section{Additional reduction}\label{sec:pipeline}
We performed an independent reduction on the TSO using \texttt{NAMELESS}; we follow the steps described in detail in \citet{coulombe2022} and Coulombe et al. (in press) to correct for instrumental effects. Bad pixels are flagged using the spatial derivative of the detector images and corrected using bicubic interpolation. The background is subtracted by independently scaling both sides of the background model provided by STScI. Cosmic ray events are addressed by computing the running median over time for all individual pixels and clipping any counts more than 4\,$\sigma$ away from their median. We then correct for 1/$f$ noise by scaling each column of the trace independently (considering only pixels within a 30-pixel distance from the center of the trace, also independently scaling orders 1 and 2) and subtracting the additive constant that minimizes the chi-squared of the fit between the column at a given integration and its median over time. Background contaminants are masked when treating the 1/$f$ noise to avoid biasing the light-curves. Finally, we extract the light-curves using a box aperture with a width of 32 pixels and use the wavelength solution from MAST. 

\begin{figure*}
	\centering
	\includegraphics[width=\textwidth]{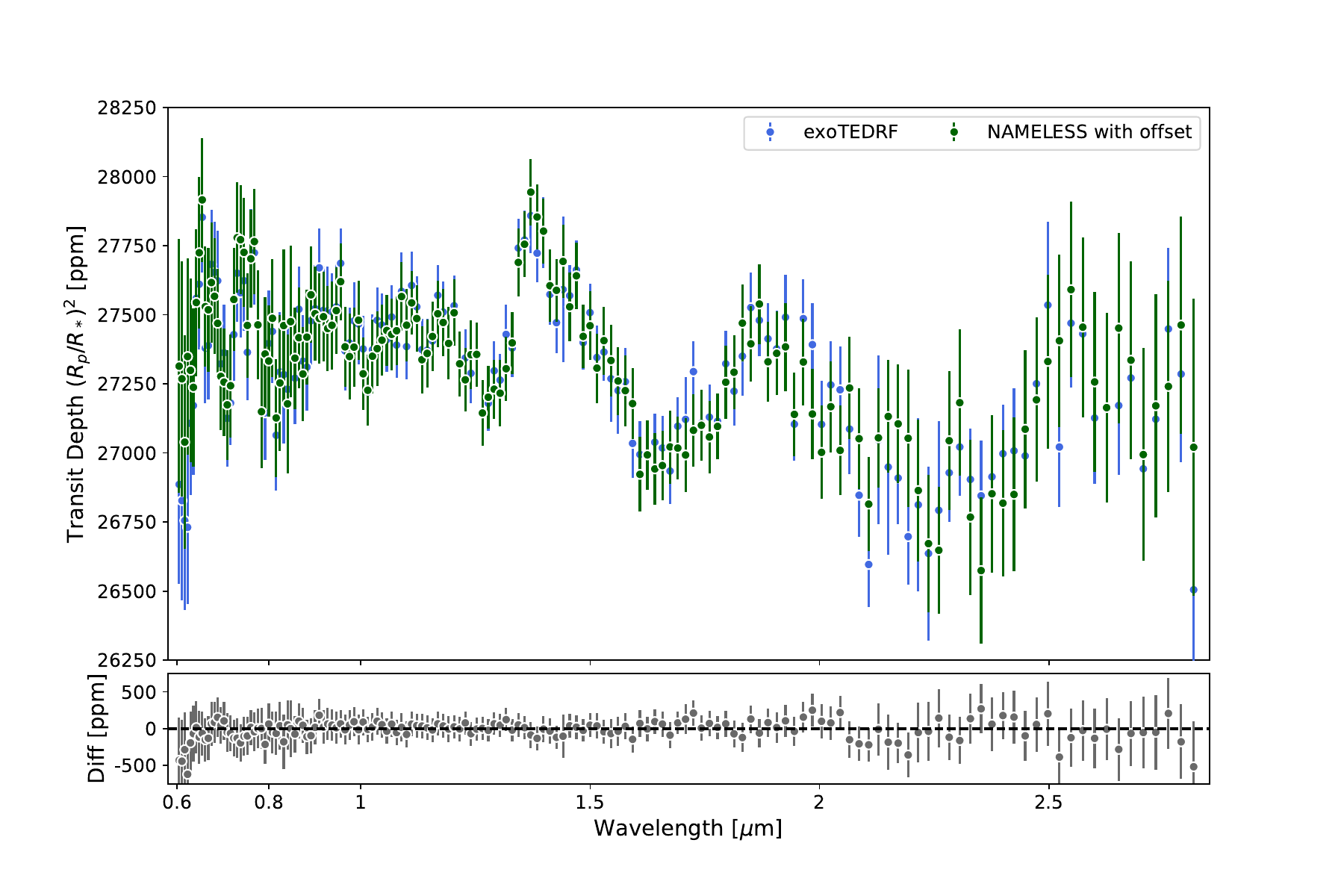}
    \caption{Comparison of NIRISS/SOSS transmission spectra obtained with two reduction pipelines: \texttt{exoTEDRF} (blue) and \texttt{NAMELESS} (green). \emph{Top:} Transmission spectra are shown binned to a resolving power of $R$ = 100. Note that an offset of -174.1\,ppm has been applied to the \texttt{NAMELESS} spectrum. \emph{Bottom}: The difference between the \texttt{exoTEDRF} and \texttt{NAMELESS} reduction is shown.
    \label{fig:ts_comparison}}
\end{figure*}

The light-curve fitting is done as described in Section \ref{sec:lc}, except that the flux errors reported by the reduction pipeline are not used; we fit for the scalar jitter term. Table \ref{tab:wlc_parameters} shows the best-fitting broadband light-curve parameters. This retrieved transmission spectrum and the one from the reference \texttt{exoTEDRF} reduction are displayed in Figure \ref{fig:ts_comparison}. These two transmission spectra are in good agreement, displaying consistent features across the entire wavelength range of NIRISS/SOSS. There is an offset of 174.1\ ppm between them, mainly due to the slight differences in the values fixed from the broadband light-curve fit.

\onecolumn\small\it\hspace{-11pt}
$^{1}$Department of Physics, University of Oxford, Parks Rd, Oxford, OX1 3PU, UK\\
$^{2}$Trottier Institute for Research on Exoplanets and Département de Physique, Université de Montréal, 1375 Avenue Thérèse-Lavoie-Roux,\\ Montréal, QC, H2V 0B3, Canada\\
$^{3}$Department of Astronomy, University of Michigan, 1085 South University Avenue, Ann Arbor, MI 48109, USA\\
$^{4}$Department of Astronomy \& Astrophysics, University of Chicago, 5640 South Ellis Avenue, Chicago, IL 60637, USA\\
$^{5}$Observatoire Astronomique, Université de Genève, 51 chemin Pegasi, 1290 Versoix, Switzerland\\
$^{6}$Department of Physics, McGill University, 3600 rue University, Montréal, QC, H3A 2T8, Canada\\
$^{7}$Department of Earth and Planetary Sciences, McGill University, 3600 rue University, Montréal, QC, H3A 2T8, Canada\\
$^{8}$Observatoire du Mont-Mégantic, Université de Montréal, Montréal, QC, H3C 3J7, Canada\\
$^{9}$NRC Herzberg Astronomy and Astrophysics, 5071 West Saanich Rd, Victoria, BC, V9E 2E7, Canada\\
$^{10}$Department of Physics and Astronomy, University of Victoria, 3800 Finnerty Rd, Victoria, BC V8P 5C2, Canada\\
$^{11}$Carl Sagan Institute and Astronomy Department, Cornell University, 302 Space Sciences Building, Ithaca, NY 14853, USA\\
$^{12}$Department of Physics and Astronomy, Johns Hopkins University, Baltimore, MD, 21218, USA

\bsp	
\label{lastpage}
\end{document}